%% file: main.tex
\documentclass[hidelinks,onefignum,onetabnum,10pt]{siamonline250211}
\usepackage[margin=0.75in]{geometry}
\usepackage{scrextend}
\changefontsizes{10pt}
\input{preamble.tex}
\usepackage[numbers,sort&compress,super]{natbib}
\setlist[itemize]{leftmargin=2em}
\setlist[enumerate]{leftmargin=2em}

\titlespacing{\subsection}{0pt}{*3}{*1.5}
\renewcommand{\thesubsection}{\thesection\alph{subsection}}
\renewcommand\thesubsubsection{\thesubsection.\roman{subsubsection}}

\usepackage{orcidlink}

\title{Direct RNA sequence design under codon constraints\\[2pt]using expressive tensor-based secondary structure models}

\author{Mark Fornace$^{1}$\,\orcidlink{0000-0002-5829-5839},
Christina Wuyan Wang$^{1,2}$\,\orcidlink{0009-0001-0446-5900}
\and Michael Lindsey$^{1,3}$\,\orcidlink{0000-0002-4508-9454} \\
$^1$Lawrence Berkeley National Laboratory; $^2$University of Colorado Boulder; $^3$University of California, Berkeley
}

\crefname{Section}{Section}{Sections}
\Crefname{Section}{Section}{Sections}
\crefname{subsection}{Section}{Sections}
\Crefname{subsection}{Section}{Sections}
\crefname{subsubsection}{Section}{Sections}
\Crefname{subsubsection}{Section}{Sections}

\begin{document}

\twocolumn
\maketitle


\begin{abstract}
       Nucleic acid sequence design via codon optimization is a fundamental task with applications across synthetic biology, mRNA therapeutics, and vaccine design. 
       Existing computational approaches seek to optimize only a single codon sequence and are limited to either a restricted class of simplified free energy models or to heuristic optimization procedures.
       In this work, we present a direct and efficient algorithm for stability-based codon optimization: we simultaneously sample many sequences from a Boltzmann distribution defined in terms of the codon sequence and a fully detailed secondary structure free energy model, together with algorithms to exactly compute statistical quantities including free energies, equilibrium base pair probabilities, and base and codon marginals.
       These draw upon a tensor-based formulation of secondary structure thermodynamics and demonstrate, for the first time, that global sequence design can be accomplished with respect to a highly accurate and systematically improvable free energy model. 
       Our algorithms leverage CPU and GPU parallelism to achieve 10-1000$\times$ speedups over state-of-the-art alternatives.
\end{abstract}

\section*{Significance}
Computational mRNA sequence design is a vital tool for current research in bioengineering, metabolic engineering, and vaccine design.
To achieve more efficient translation and resistance to degradation, common approaches seek a single RNA sequence that optimizes a stability objective under amino acid coding constraints. 
However, existing approaches are limited by either simplistic free energy models or heuristic optimization methods.
We develop the first direct algorithm to sample an arbitrary number of stable RNA sequences, rather than just one, using an exact sampling procedure and a highly accurate and systematically improvable tensor-based free energy model.
CPU and GPU parallelism yields massive acceleration within our framework, which is dramatically faster than leading alternatives.

\section{Introduction}

The computational design of an mRNA sequence encoding a target protein is a primary goal of current research in computational biology and bioengineering.
Efficient mRNA translation is key to many applications, including 
    synthetic biology \cite{Schmidt2023-maximizing,Coleman2008-virus,Mali2013-rna-guided,Lajoie2013-genomically},
    vaccine development \cite{Vogel2021-bnt162b,Kim2024-computational,Larocca2016-vaccine,Jin2025-mRNA}, and 
    nucleic acid therapies \cite{Paddon2013-high-level,Kormann2011-expression}.
Since multiple codons (i.e., triplets of RNA bases) can code for the same amino acid, the number of mRNA sequences coding for the same protein grows exponentially with the number of amino acids. 
The vastness of this space grants us considerable freedom to optimize the RNA sequence's fitness according to criteria beyond simple codon constraints. 

Beyond the basic (and inviolable) constraints of amino acid coding, a large body of research (e.g., \cite{Gustafsson2004-codon,Wayment-Steele2021-theoretical}) has posited various soft constraints depending on other hypothesized principles of mRNA fitness.
Of particular importance and ubiquity is the principle of \emph{stability maximization}, which suggests that the secondary structure (i.e., discrete base pairing state) of the mRNA coding sequence should be predominantly base-paired, in order to enhance translation efficiency \cite{Kim2022-modifications,Wayment-Steele2021-theoretical} and avoid degradation by RNases and other biological factors \cite{Wayment-Steele2021-theoretical,Mao2014-deciphering,Mauger2019-mrna,Ringner2005-folding}. 
As base pairing is highly correlated with secondary structure free energy, minimization of the mRNA sequence free energy is a primary objective for respective algorithms. 

Alongside the \emph{structural} objective of stability maximization, a number of purely \emph{sequence-based} objectives have also been formulated based on empirical observations of coding efficiency.
In formulations like the codon adaptation index (CAI) \cite{Sharp1987} or codon pair bias (CPB) \cite{Coleman2008-virus}, codons are chosen with an aim towards reproducing codon or codon pair frequencies observed in a target genome.
Other design principles operate similarly, penalizing or bonusing the optimized RNA sequences based on local subsequence information \cite{Condon2012,Gustafsson2004-codon}.
These considerations motivate an algorithm to perform stability maximization under the hard constraints of amino acid coding and the soft constraints of local sequence-based objectives.

\subsection*{Contributions}
In this work, we develop a framework for sequence optimization purely based on a combination of dynamic programming with tensor-based models of secondary structure and codon constraints.
We demonstrate:

\newcommand{\LessPrefactor}[1]{%
  \ensuremath{%
    \mathrel{\raise0.3ex\hbox{$\scriptstyle\lesssim$}}%
    #1\times%
  }%
}
\begin{itemize}[itemsep=0pt]
    \item principled and direct stability-based design using dynamic programming with \emph{fully detailed} secondary structure free energy models
    \item an \emph{exact and principled sampling procedure}. 
    Rather than producing just one sequence, our approach defines an entire \emph{ensemble} of coding sequences which may be considered for downstream use.
    \item new methods for \emph{exactly computing marginal probabilities} over our sampling distribution, including base pair probabilities and codon frequencies
    \item a novel \emph{tensor train formulation} of hard and soft constraints, extensible to a broad class of sequence-based objectives
    \item \emph{efficient GPU-accelerated algorithms}, which incur just a cost prefactor of \LessPrefactor{\ObservedPrefactor} over RNA sequence analysis.
    A 5000 amino acid design takes  $\approx 7$ minutes.
    Our benchmarks show 10-1000$\times$ speedups over previous algorithms.
\end{itemize}

In sum, our approach demonstrates, for the first time, that stability maximization may be solved via dynamic programming for a realistic secondary structure free energy model.
Uniquely, we show that our main sampling-based methodology offers (1) an efficient generator of many sequences, which may be chosen or filtered according to arbitrary experimentalist preferences afterwards, (2) what we expect is empirically a good compromise between entropic and enthalpic criteria, avoiding pathological sequences in favor of those drawn from a statistically principled ensemble.
The methodology thus offers a general and extensible framework for sequence design via dynamic programming.






\subsection*{Alternative methods}


Owing to the ubiquity of the codon optimization problem, a number of computational methods have been previously proposed.
We categorize common approaches here%
\footnote{See the reviews of \cite{Ward2025-mrna,demissie2025comparative,Jin2025-mRNA,Kim2024-computational} for additional methods and details.}.
To our knowledge, there has been no perspective that unifies the problem of codon optimization, dynamic programming, and sequence sampling as we do. 


\paragraph*{Dynamic programming and related approaches}
Past dynamic programming methods have generally yielded algorithms based on minimum free energy (MFE) minimization using simplified free energy models \cite{Terai2015,Cohen2003-NaturalSelection,Gu2024DERNA}.
Recent high-profile work includes applications of lattice parsing to simultaneously optimize MFE and CAI objectives \cite{Zhang2023-algorithm} and probabilistic lattice parsing to optimize ensemble free energy \cite{Dai2025-EnsembleDesign}.
Performance benchmarks of selected algorithms demonstrate dramatic speedups and are compared in \cref{si:alternative-methods-comparisons}.

\paragraph*{Approaches based on local search}
Algorithms based on local search can flexibly optimize sequences with respect to diverse objectives, including stability. 
Such methods commonly begin with an arbitrary sequence and mutate a single codon at a time, re-evaluating the objective and accepting changes which improve the objective \cite{Zulkower2020-dna,Vostrosablin2024-mRNAid,Fornace2022-nupack,Zadeh2011-nucleic,Wolfe2017-constrained,Chung2012-computational}.
In principle, such hill-climbing methods can also be extended to sampling via Markov Chain Monte Carlo (MCMC) to avoid being trapped in local minima. 
However, objective re-computation can be costly, and the number of iterations for convergence is not bounded \emph{a priori}.

\paragraph*{Deep learning approaches}
Recently, machine learning methods based on large language models (LLMs), transformer-based models, and other generative models have been increasingly applied to codon optimization \cite{Constant2023-deep,faizi2025,Fallahpour2025,Sidi2025-predicting,HeZhang2025,Li2025-RiboDecode,Li2025-ARCADE,Ren2024-CodonBERT,Elazar2025-interrogating}. 
Such algorithms design \\ host-specific sequences by learning codon usage patterns across organisms in a flexible, albeit black-box, manner.
Downsides include a limited amount of translation efficiency data for training and a lack of interpretability.







\subsection*{Mathematical notation}

Following on past work, we commonly use zero-based semi-inclusive indexing, such that $\Range{a}{b} \Eq a,\dots,b-1$ with the convention that $\Range{a}{b}$ is an empty sequence if $a=b$.
We use the shorthand $\UpTo{n} \Eq \Range{0}{n}$ and denote the length of any sequence $s$ as $\Abs{s}$.
We commonly assume cyclic indexing such that, e.g., $\phi_{n+1} = \phi_{1}$.
For a secondary structure of specified RNA sequence $\phi$, we say that $i \cdot j$ if $i \neq j$ and the nucleotides at sequence positions $i$ and $j$ are base paired, or if $i=j$ and the nucleotide at position $i$ is unpaired.
In tensor notation, we reserve subscripts for sequence position indices and superscripts for all other indices. $\Id$ is the identity matrix. A full glossary of notation is provided in \cref{si:notation}.

\begin{figure}[!tbp] \input{./floats/secondary-structures.tex} \end{figure}

\section{Preliminaries and contributions}

In this section, we lay out our problem statement, notation, and contributions. 
We begin by precisely defining our considered ensembles and probability distributions.

\subsection{Sequence and structural ensembles}
Given a target amino acid sequence $\psi$ consisting of individual amino acids $\psi_i \in \{\texttt{A}, \texttt{R}, \texttt{N}, \dots\}$, let $\Phi(\psi)$ be the set of all RNA sequences $\phi$ consisting of individual bases $\phi_i \in \{\texttt{A}, \texttt{C}, \texttt{G}, \texttt{U}\}$, which code for $\psi$.
This means that each codon (word of 3 consecutive bases) within $\phi$ comes from a fixed library of codons which code for the corresponding amino acid in the target sequence $\psi$.
For instance, \texttt{D} (aspartic acid) at position $i$ in amino acid sequence $\psi$ would necessitate that the RNA subsequence $\phi_{\Range{3i}{3i+3}}$ must be either \texttt{GAU} or \texttt{GAC}.
For simplicity, we omit noncoding regions from our description (\cref{si:noncoding}), so that $\Abs{\phi} = 3 \Abs{\psi}$.
We will commonly use the notation $n \Eq \Abs{\phi}$.

For a fixed RNA sequence $\phi$, we consider the ensemble $\Gamma(\phi)$ of all unpseudoknotted secondary structures $s$ compatible with that sequence, as in past work, e.g., \cite{Fornace2020-unified}.
An unpseudoknotted secondary structure (for a single strand\footnote{For simplicity, we restrict ourselves to RNA and amino acid sequences involving only one strand, as this is sufficient for most codon optimization problems of practical interest.}) is one which has no crossing base pairs when plotted in the form of \cref{fig:structures}a, i.e., in which no two base pairs $i \cdot j$ and $k \cdot l$ (ordered so that $i<j$ and $k < l$) satisfy $i < k < j < l$.
The unpseudoknottedness assumption is well-studied 
and enables exact dynamic programs with $O(n^3)$ computational complexity.

Finally, for a given secondary structure, it is an implicit feature of the tensor models introduced in \cite{Fornace2025-new} that a \emph{subensemble} of hidden configurations underlies a given secondary structure.
That is, a secondary structure is compatible with one or more configurations, each of which corresponds to a set of integers specifying each of the indices in the respective tensor diagram for the secondary structure partition function.
For a given secondary structure $s$, we refer to this subensemble as $\StateEnsemble(s)$ and each configuration therein as $\zeta$.

\subsection{Probability distributions}
Given the three ensembles $\SeqEnsemble$, $\StructureEnsemble$, and $\StateEnsemble$, we define the probability distribution
$\Dist(\psi)$ with probability mass function $\Prob(\phi,s,\zeta) \propto e^{-\logq(\phi,s,\zeta)}$ for $\phi \in \SeqEnsemble(\psi), s \in \Gamma(\phi), \zeta \in \StateEnsemble(s)$, where $\logq(\phi,s,\zeta)$ is the unitless Gibbs free energy of state $(\phi, s, \zeta)$.
We formally extend $\Dist(\psi)$ to have zero probability mass outside of $\Phi(\psi)$, covering the space of \emph{all} RNA sequences of length $3 \Abs{\psi}$:
\begin{eqn}
    \qpsi(\phi,s,\zeta) \Eq \begin{cases} 
        \logq(\phi,s,\zeta) & \phi \in \Phi(\psi) \\
        \infty & \text{otherwise}
    \end{cases},
\end{eqn}
Important statistics may be calculated by marginalizing over $\Dist(\psi)$ in various ways. 
Specifically, we obtain the free energy of a given RNA sequence $\phi$ and secondary structure $s$ by marginalizing over all compatible configurations $\zeta$:
\begin{eqn} \label{eq:structure-q}
    \qpsi(\phi,s) &\Eq -\log \sum_{\zeta \in \StateEnsemble(s)} e^{-\qpsi(\phi, s, \zeta)}, 
\end{eqn}
the free energy of a given RNA sequence $\phi$ by marginalizing over all compatible secondary structures $s$:
\begin{eqn} \label{eq:sequence-q}
    \qpsi(\phi) &\Eq -\log \sum_{s \in \Gamma(\phi)} e^{-\qpsi(\phi, s)},
\end{eqn}
and the free energy associated to a given amino acid sequence $\psi$ by marginalizing over all compatible RNA sequences $\phi$:
\begin{eqn} \label{eq:amino-q}
    \qpsi &\Eq -\log \sum_{\phi \in \SeqEnsemble(\psi)} e^{-\qpsi(\phi)}.
\end{eqn}
As such, $\qpsi(\phi,s)$ and $\qpsi(\phi)$ are precisely the physical Gibbs free energies of a nucleic acid secondary structure $s$ and complex $\phi$, while $\qpsi$ represents the Gibbs free energy over all compatible RNA sequences.

Finally, we generalize our formulation to handle common RNA sequence-based objectives.
Incorporating these objectives via a biasing function $\Bias_\psi(\phi)$, we define the probability distribution $\Dist^\Bias(\psi)$ on the same space as $\Dist(\psi)$ with probability mass function $\Prob^\Bias(\phi,s,\zeta) \propto e^{-\logq(\phi,s,\zeta) - \Bias_\psi(\phi)}$.
This modification by $\Bias_\psi(\phi)$ can be handled without affecting our essential algorithmic structures, so we will focus on $\Dist(\psi)$ in our presentation and point out the differences only when relevant.


\subsection{Computed quantities\label{s:quantities}}
Our new algorithms (1) calculate $\qpsi$ exactly and (2) directly sample RNA sequences $\phi$ from $\Dist(\psi)$ and more generally $\Dist^\Bias(\psi)$.
These algorithms have $O(n^3)$ complexity, invoke no approximations, and incur only a roughly \LessPrefactor{\ObservedPrefactor} cost prefactor compared to analysis of a fixed RNA sequence.

Moreover, we demonstrate how to exactly compute certain essential statistics of the sampling distribution $\Dist(\psi)$.

First, consider the equilibrium base pair probability matrix $\Ppsi$ (for each $i\in \UpTo{n},j \in \UpTo{n}$) defined entrywise by:
\begin{eqn} \label{eq:pair-probability}
    \Ppsi_{i, j} = \Prob\lrp{i \cdot j \mid s \sim \Dist(\psi)},
\end{eqn}
generalizing the convention for a fixed nucleic acid sequence $\phi$ \cite{Fornace2020-unified}. 
Given this matrix, it is straightforward to compute the expected total number of base pairs in a sampled sequence at thermal equilibrium.
We show how to compute the full matrix $\Ppsi$ exactly with $O(n^3)$ cost.

If one samples a sequence $\phi$ from $\Dist(\psi)$, what is the probability that $\phi_{\Range{3i}{3i+3}}$ corresponds to codon $c$?
Moreover, what is the probability that base $b$ appears at $\phi_i$?
If one can answer these questions, for example, it is simple to compute the expected total number of occurrences for any codon or base.
Specifically, consider the marginal probability of nucleotide $\phi_i$ being a given base $b$ (e.g., $b=\texttt{G}$):
\begin{eqn} \label{eq:marginal-base}
    \Prob_i^\phi(b) = \Prob\lrp{\phi_i = b \mid \phi \sim \Dist(\psi)},
\end{eqn}
and the marginal probability of amino acid $\psi_i$ being coded by a given codon $c$ (e.g., $c=\texttt{GAU}$) as:
\begin{eqn} \label{eq:marginal-codon}
    \Prob_i^\psi(c) = \Prob\lrp{\phi_{\Range{3i}{3i+3}} = c \mid \phi \sim \Dist(\psi)}.
\end{eqn}
For a given $i$, we show how to compute both $\Prob_i^\phi(b)$ and $\Prob_i^\psi(c)$ exactly with $O(n)$ complexity.
This implies an additional $O(n^2)$ cost to compute all such marginals, after an initial $O(n^3)$ computation.

Finally, we demonstrate that by application of an evaluation algebra approach \cite{Fornace2020-unified}, we can exactly solve the minimization problem 
\begin{eqn} \label{eq:min-problem}
       \qmin \Eq \min_{\phi,s,\zeta} \ \left\{ \qpsi(\phi, s, \zeta) + \Bias_\psi(\phi) \right\}
\end{eqn}
over a suitably defined ensemble of sequence $\phi$, secondary structure $s$, and configuration $\zeta$.
In fact, this same evaluation algebra approach also yields our algorithms for sequence sampling.

In contrast, while we do not exactly solve the related minimization problem 
\begin{eqn}
       \min_{\phi} \ \left\{ \qpsi(\phi) + \Bias_\psi(\phi) \right\},
\end{eqn}
we show that our solution of \cref{eq:min-problem} yields \emph{approximate} solutions for this problem as well.
In general, we view the sampling problem as more fundamental than this minimization problem, for both practical and theoretical reasons, as we will discuss in \cref{s:discussion}.



\begin{figure}
    \centering
    \includegraphics[width=.95\linewidth]{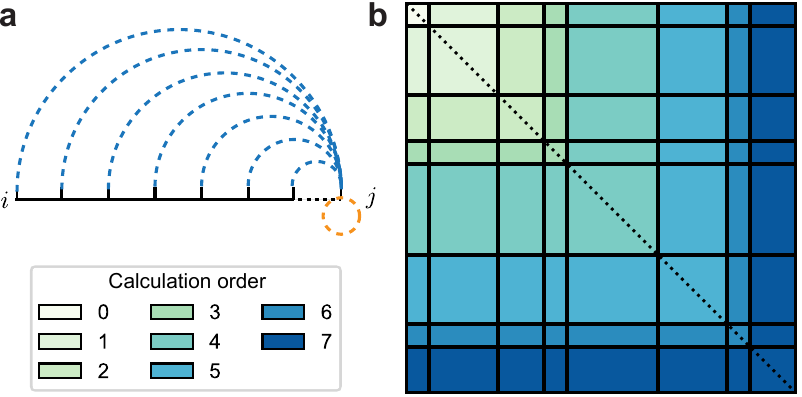}
    \caption{\label{fig:structures}
        \Bullet{a} Conceptual idea of partition function dynamic programs, in which an additional 3$'$-most base $j$ is added and incorporated.
        That base may remain unpaired (orange), or it may close a base pair $i \cdot j$ originating from any earlier base $i < j$ (blue).
        Considering all possibilities incurs an $O(n)$ summation for the recursion.
        Recursing through all $O(n^2)$ subsequences gives an overall $O(n^3)$ cost.
        \Bullet{b} Block matrix layout implied by our dynamic programs for sequence sampling.
        Each block $Q_{i,j}$ is the partition function matrix of subsequence $\Range{i}{j}$ if $i \leq j$ or else the partition function matrix of $\Range{j}{i}$ constrained such that bases $j$ and $i-1$ are paired (or unpaired, in the case of $j=i-1$).
        Compared to the simpler problem of sequence analysis, in which each recursion block $Q_{i,j}$ is square \cite{Fornace2025-new}, the different dimensionalities arising from codon constraints (\cref{fig:transfer}) imply non-square blocks. 
    }
\end{figure}

\section{Free energy computation \label{s:free-energy}}

In this section, we tackle the core problem of computing $\qpsi$ using dynamic programming.
This represents the free energy (or negative log partition function) obtained by summing over all possible RNA sequences, secondary structures, and configurations compatible with amino acid sequence $\psi$.
Related algorithms for sampling and minimization (\cref{s:downstream}) can be readily derived using the evaluation algebra concept of \cite{Fornace2020-unified}.
Therefore, this section describes the core intellectual leap that enables the treatment of the generalized sequence sampling problem.



As mentioned earlier, our algorithms draw heavily upon the tensor modeling paradigm of \cite{Fornace2025-new}.
In \cref{s:tensor-constraints}, therefore, we phrase the codon constraint satisfaction problem in the same language, describing how such sequence constraints can be formulated in terms of \emph{tensor trains} \cite{oseledets2011tensor}.
Then, in \cref{s:structure-dp,s:sequence-dp}, we describe how the secondary structure and sequence tensor diagrams can be combined, yielding a \emph{folding} function $\DesignFold$ which generalizes the folding function $\Fold$, which was defined in \cite{Fornace2025-new} for a fixed RNA sequence.
We show how this substitution may be performed cleanly \emph{without changing} the main free energy algorithm of \cite{Fornace2025-new} at all, necessitating only a mild generalization of our high-performance computing implementation.

\subsection{Formulating codon constraints using tensor trains\label{s:tensor-constraints}}

We first show how the hard and soft constraints of codon optimization may be formulated using the concepts of tensor trains and, relatedly, traces of matrix products.
Within this context, we can ignore any consideration of secondary structure, focusing only on RNA sequence information.
Consequently, our secondary structure algorithms are independent of the details of the tensor trains specified here. As such, we focus now on defining an ensemble over RNA sequences $\phi$ defined only by hard codon constraints and a soft biasing function $\Bias_\psi(\phi)$. 

\begin{figure*}
    \vspace{0.5em}
       \centering
       \includegraphics[width=.82\linewidth]{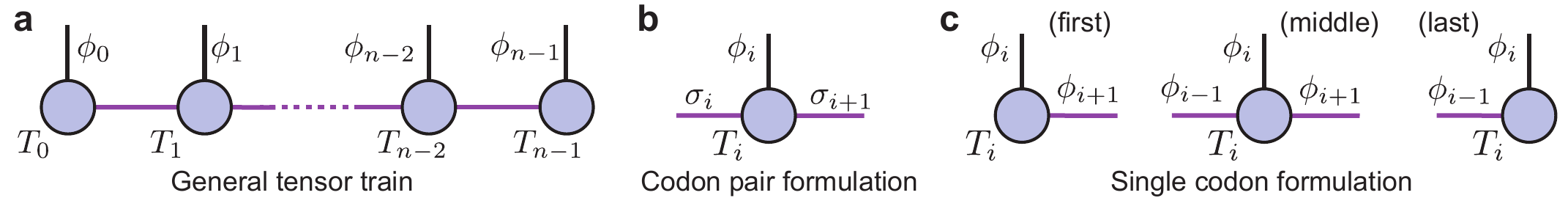}
       \caption{Tensor train formulation of codon constraints.
       \Bullet{a} 
       Without any consideration of secondary structure, hard and soft sequence constraints can be implemented using a tensor train. 
       \Bullet{b} Depiction of an individual tensor core $T_i$ in the codon pair formulation.
       The tensor is indexed by the base $\phi_i$, codon index $\sigma_i$ of the base $\phi_{i-1}$, and codon index $\sigma_{i+1}$ of the base $\phi_{i+1}$, yielding a nonzero element only if the combination is possible. 
       \Bullet{c}
       In the single codon formulation, the left and right tensor cores for each codon are given by the identity matrix.
       The middle tensor is indexed by $\phi_{i-1}$, $\phi_i$, and $\phi_{i+1}$, yielding a nonzero element only if that triplet comprises a valid codon. 
       }
       \label{fig:transfer}
\end{figure*}

To wit, define a probability distribution $\SttDist(\psi)$ over all length-$n$ RNA sequences $\phi$ with probability mass function proportional to $e^{-\qseq_\psi(\phi)}$, where:
\begin{eqn}
       \qseq_\psi(\phi) \Eq -\log \Tr\lrs{T_{0}^{\phi_0} \cdots T_{n-1}^{\phi_{n-1}}}.
\end{eqn}
Here, we define each $T_{i}^{\phi_i}$ to be a matrix of shape $(\DimTT_i,\DimTT_{i+1})$, and, therefore, each $T_{i}$ can be viewed as a 3-index tensor.
The probability mass function for $\SttDist(\psi)$ is then precisely a \emph{tensor train} \cite{oseledets2011tensor} (\cref{fig:transfer}a), which we call the \emph{RNA sequence tensor train}. 
Denoting $\bar{T}_i = \sum_{\phi_i} T_{i}^{\phi_i}$, we may marginalize over all RNA sequences $\phi$ to obtain the free energy of this ensemble as:
\begin{eqn}
       \qseq_\psi \Eq -\log \Tr\lrs{\bar T_{0} \cdots \bar T_{n-1}},
\end{eqn}
such that 
the partition function of the ensemble is simply the trace of a product of transfer matrices.

\hspace{-1em} Next, we will demonstrate how to choose $(T_0, \dots, T_{n-1})$ so that the distribution $\SttDist(\psi)$ precisely implements our desired hard and soft codon constraints.


\subsubsection{Incorporating hard constraints}

We first design a simple tensor train such that $\qseq_\psi(\phi)$ is 0 if $\phi \in \Phi(\psi)$, else $\infty$.
In such a case, $e^{-\qseq_\psi} = \Abs{\Phi(\psi)}$, i.e., $\qseq_\psi$ is analogous to counting the number of possible RNA sequences $\phi$ obeying $\psi$'s coding constraints.
Using the Kronecker $\delta$ and modular arithmetic, define: 
\newcommand{\Compatible}{\mathcal{C}}
\begin{eqn}
\label{eq:pair-tt} 
    T_{i}^{\phi_i, \sigma_a, \sigma_b} = \Compatible_\psi(\phi_i, \sigma_b) \times  \begin{cases}
        \delta_{\sigma_a,\sigma_b}
         & i \bmod 3 = 0 \\
        \delta_{\sigma_a,\sigma_b}  & i \bmod 3 = 1
        \\
        1
         & i \bmod 3 = 2
    \end{cases},
\end{eqn}
where $\sigma_a,\sigma_b$ correspond to codon indices and $\Compatible_\psi(\phi_i, \sigma_b)$ returns 1 if base $\phi_i$ is present at the corresponding position in codon $\sigma_b$, else 0.
We observe that the dimensionality of $T_i$ in this specification is (\#possible bases at $i$, \#possible codons at $i-1$, \#possible codons at $i$) and visualize it in \cref{fig:transfer}b.
Since \cref{eq:pair-tt} generally involves pairs of neighboring codons, we call it the \emph{codon pair} tensor train for $\psi$.

In fact, we can construct a leaner tensor train by omitting the redundant information in \cref{eq:pair-tt}: 
\begin{eqn}
\label{eq:single-tt} 
    T_{i}^{\phi_i, \sigma_a, \sigma_b} = \begin{cases}
        \delta_{\phi_i,\sigma_b}
         & i \bmod 3 = 0 \\
        \Compatible_{\psi}'(\sigma_a, \phi_i, \sigma_b) & i \bmod 3 = 1
        \\
        \delta_{\phi_i,\sigma_a}
         & i \bmod 3 = 2
    \end{cases},
\end{eqn}
where $\sigma_a,\sigma_b$ now denote \emph{base} indices and $\Compatible'_\psi(\sigma_a, \phi_i, \sigma_b)$ returns 1 if $(\sigma_a, \phi_i, \sigma_b)$ is a valid codon for $\psi$ at bases $(i-1,i,i+1)$.
Since \cref{eq:single-tt} involves only a single codon, we call it the \emph{single codon} tensor train for $\psi$.
We observe that the structure of \cref{eq:single-tt} yields different dimensions for $T_i$ depending on the position of $i$ within its respective codon (\cref{fig:transfer}c).

\subsubsection{Incorporating soft constraints}

Both the single codon and codon pair tensor trains encode the hard constraint of amino acid coding.
That is, computation of the partition function of $\SttDist(\psi)$  under either choice simply counts the RNA sequences in $\Phi(\psi)$, yielding $\qseq(\psi) = -\log \,\Abs{\Phi(\psi)}$.
However, observe that it is easy to modify either tensor train to incorporate sequence-dependent biasing factors, in order to guide sequence design via user-chosen soft constraints.

Specifically, consider the simplest case in which a specific codon $c$ is penalized by free energy $f_c$ at amino acid position $i$ (as in CAI, for example).
In the single codon tensor train, scaling $T_{3i+1}^{\phi_{3i+1},\sigma_a,\sigma_b}$ by $e^{-f}$ yields that $\qseq_\psi(\phi) = f \cdot \delta(\text{codon } i = (\sigma_a, \phi_{3i+1}, \sigma_b) \text{ in } \phi)$ for any $\phi \in \Phi(\psi)$.
Thus, the single codon tensor train allows for any simple soft constraints depending on a single codon in isolation.

Correspondingly, the codon pair tensor train allows for any soft constraints depending on a pair of adjacent codons.
Namely, to apply a free energy penalty of $f$ for neighboring codons $\sigma_a$ and $\sigma_b$ at positions $i-1$ and $i$, respectively, we scale $T_{3i}^{\phi_{3i},\sigma_a,\sigma_b}$ by $e^{-f}$.
Thus, the codon pair tensor train is more flexible than the single codon tensor train, but its increased dimensionalities incur more computational expense.

For more complicated soft constraints (e.g., those for motif search), more information must be embedded, increasing the tensor train dimensionalities and downstream cost.
In \cref{si:operators,si:codon-cost}, we discuss (1) a general formulation for sequence tensor train construction and (2) a simple analysis of the incurred computational costs.
For common systems, remarkably, we demonstrate that single codon and codon pair tensor trains incur only theoretical prefactors of \LessPrefactor{7} and \LessPrefactor{50} over the cost of fixed sequence free energy computation, respectively.
These prefactors are even lower in practice (\cref{s:performance}).




\subsection{Dynamic programs over structures for a given sequence\label{s:structure-dp}}

For a given RNA sequence, the tensor-based model of \cite{Fornace2025-new} expresses the free energy of a secondary structure via the contraction of a tensor diagram with a topology induced by that secondary structure (\cref{fig:model}cd).
Meanwhile, in \cref{s:tensor-constraints} we showed how the hard and soft constraints of codon optimization may be encoded into a tensor train which specifies a distribution $\SttDist(\psi)$ of RNA sequences $\phi$ given an amino acid sequence $\psi$.
We now describe how to combine these two tensor approaches cleanly into a free energy algorithm summing concurrently over all RNA sequences and secondary structures.


Before describing our new approach, we review the essential elements of the main free energy algorithm of \cite{Fornace2025-new}, which sums over all secondary structures compatible with a \emph{fixed} RNA sequence.
In that methodology, $Q$ is a single square matrix laid out in blocks of size $(\Dim, \Dim)$ and computed over the course of the algorithm, with $\Dim$ denoting the dimensionality of the free energy model. 
Free energy computation proceeds by the \emph{single} dynamic programming recursion:
\begin{eqn}
    Q_{i, j} &= \begin{cases}
        \sum_{k \in \Range{i}{j}} Q_{i,k} Q^\t_{j,k} & i < j \\
        \Fold_{j,i}(Q_{j+1,i-1}) & \text{otherwise}
    \end{cases} \label{eq:dp-short},
\end{eqn}
in which each element $Q_{i,j}$ is a block of the matrix $Q$ of dimension $(\Dim, \Dim)$ such that the expression $Q_{i,k} Q^\t_{j,k}$ is a multiplication of two $(\Dim, \Dim)$ matrices, the second being transposed.
Central to this methodology is the $\Fold$ folding function, which for a \emph{fixed} RNA sequence $\phi$ is defined as:
\begin{eqn} \label{eq:fold}
    \Fold_{i,j+1}(X) &\Eq \begin{cases}
        \Id & i = j+1 \\
        \V_{i}^\t & i = j \\
        \sum_{\Pp \in \UpTo{\Np}} B_{i,j}^{\Pp,\t} \TrS[B_{j,i}^{\Pp} X] & \text{otherwise}
    \end{cases},
\end{eqn}
where the dependence of $V_i$ and $B_{i,j}^\rho$ on the
bases $\phi_i$ and $\phi_j$ is omitted from the notation, $\rho$ indexes over possible base pair conformations in the range $\UpTo{\Np}$, $\Np$ being a model parameter, and $\TrS[X] \Eq \Tr[S X]$ for a matrix $S$ which is also a model parameter. 
Importantly, the input RNA sequence $\phi$ and free energy model parameters affect the recursion \cref{eq:dp-short} only through the $\Fold$ function.

\begin{figure}[!b] \input{./floats/tensor-contraction.tex} \end{figure}

Given this function, a dynamic program can evaluate the recursive formula \cref{eq:dp-short} by means of a successively growing sequence (\cref{fig:structures}b).
After computing all $n^2$ blocks of $Q$, the final free energy of the RNA sequence $\phi$ can be computed as:
\begin{eqn} 
    \logq(\phi) = -\log \sum_{k \in \UpTo{n}} \TrS[Q_{0,k} \, \Fold_{k,\Nphi}^\t(Q_{k+1,\Nphi-1})].
    \label{eq:complex-logq}
\end{eqn}
We will next demonstrate how \cref{eq:dp-short,eq:fold,eq:complex-logq} can be generalized to incorporate summation over sequences.

\begin{figure}
       \centering
       \includegraphics[width=\columnwidth]{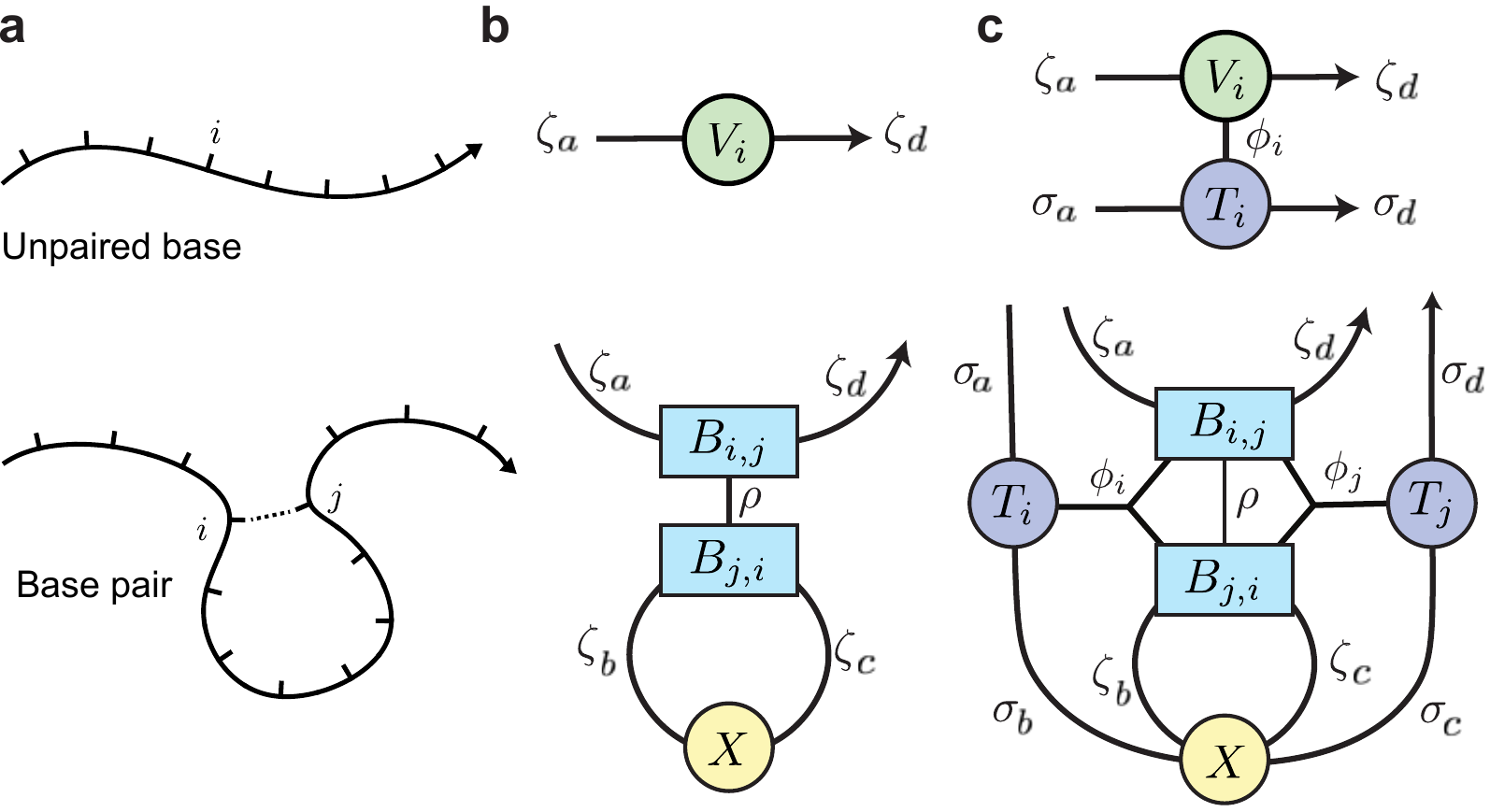} 
       \caption{\label{fig:method}
       Schematics for the folding functions $\Fold$ and $\DesignFold$. 
       Depiction of the $\Fold$ operation and equivalent tensor network contractions for unpaired base (top) and base pair (bottom) incorporation into dynamic programming recursions, in multiple settings of increasing complexity.
       \Bullet{a} Depiction of the considered operations, in which a 3$'$-most unpaired or paired base is considered.
       \Bullet{b} Existing tensor network formulation for a fixed sequence $\phi$. Here, $\Fold$ is defined just to close the bottom loop via matrix product traces and incorporate the results into the top with correlations modeled by $\rho$.
       \Bullet{c} Tensor network formulation summing over RNA sequences satisfying locally determined hard constraints, as in codon optimization. 
       $T$ tensors are introduced to propagate those constraints along newly inserted degrees of freedom.
       }
       \vspace{-2em}
\end{figure}

\subsection{Dynamic programs over sequences and structures\label{s:sequence-dp}}


We first consider the problem of summing over all possible sequences for a \emph{single} example secondary structure.
\cref{fig:contraction} illustrates two cases.
First, we lay out the RNA sequence tensor train (\cref{s:tensor-constraints}) to follow the backbone of the structure diagram.
Then we connect respective edges between the RNA sequence tensor train and the structural tensor diagram, implying a summation over all possible sequences $\phi$.
Contracting these new edges, we obtain a tensor diagram with the same topology as our original structure diagram, in which the dimensionalities of the internal edges have been changed.



It follows that we can use the \emph{same} dynamic programming framework outlined in \cref{s:structure-dp}, by replacing the $\Fold$ function with a generalized version to accomplish the additional sequence summations.
In full detail, we construct an expanded folding function $\DesignFold$ (\cref{fig:method}) as follows.
 Using $\sigma$ and $\zeta$ to represent sequence and structural summation indices, respectively, we derive:
 \vspace{-1.2em}
\begin{eqn} \label{eq:design-fold}
    \DesignFold_{i, j+1}^{\sigma_d, \zeta_d, \sigma_a, \zeta_a}(X) \hspace{-0.2em} \Eq 
    \begin{cases}
    \mathbb{I} & \hspace{-6.1em}  i = j+1 \vspace{0.1em} \\[0.4em] 
         \begin{aligned}  \sum_{\substack{\phi_i',\\\sigma_b,\sigma_c}} \hspace{0.3em} &T_i^{\phi_i',\sigma_a, \sigma_d}  (\Fold |_{\phi_i = \phi_i'})_{i,j}^{\zeta_d, \zeta_a}(X^{\sigma_b, \sigma_c}) \end{aligned} & \hspace{-5.5em}  \\[-1.5em]
          & \hspace{-6.1em} i = j \\[0.4em]
        \begin{aligned} \sum_{\substack{\phi_i', \phi_j'\\\sigma_b,\sigma_c}} \hspace{-0.2em} &T_i^{\phi_i', \sigma_a, \sigma_b} T_j^{\phi_j',\sigma_c, \sigma_d} \\[-1.75em] & \hspace{0.6em}(\Fold |_{\phi_i = \phi_i',\phi_j = \phi_j'})_{i,j+1}^{\zeta_d, \zeta_a}(X^{\sigma_b, \sigma_c}) \end{aligned} & \hspace{-6.2em}  \begin{aligned}\text{otherwise}, \\[1.4em] \end{aligned}
    \end{cases}
\end{eqn}
where the notation indicates that invocations of $\Fold_{i,j}$ are conditional on the assignments of $\phi_i$ and $\phi_{j-1}$. 

As seen in \cref{eq:design-fold}, $\DesignFold$ reflects a summation of $\Fold$ over each base possibility involved in the contraction ($\phi_i$ for an unpaired base, and $\phi_i$, $\phi_{j}$ for a base pair), using $\Fold$ to handle the structural $\zeta$ indices.
Throughout the summation, hard and soft RNA sequence constraints are contributed by the respective $T$ tensors along the $\sigma$ indices.

To incorporate the additional dimensionalities in $\DesignFold$ vs $\Fold$ (\cref{fig:method}b vs \cref{fig:method}c), finally, we expand the definition of our basic recursion quantity $Q_{i,j}$.
Specifically, while each block $Q_{i,j}$ is of size $(\gamma, \gamma)$ for a \emph{fixed} RNA sequence, allowing for a \emph{variable} RNA sequence with codon constraints implies that each block $Q_{i,j}$ is itself a tensor of size $(\gamma,\DimTT_i,\gamma,\DimTT_j)$.
Importantly, though, the conceptual distinction between sequence ($\sigma$) and structure ($\zeta$) degrees of freedom is only necessary within the fold operation (i.e., inside of the $\Fold$ function).
To the broader dynamic programming algorithm, these degrees of freedom may be combined, such that each $Q_{i,j}$ block is treated as a \emph{rectangular} $(\gamma \cdot \DimTT_i,\gamma \cdot \DimTT_j)$ matrix and all other algorithms are \emph{entirely unchanged}.
Namely, after one replaces the $\Fold$ operation as described, recursions \cref{eq:complex-logq} and \cref{eq:dp-short} can be generically invoked to compute $\qpsi$ without modification. 
\Cref{alg:complex-pf,alg:fast-pf} in \cref{si:pf} give full pseudocode for the implied approach.


\section{Additional algorithms \label{s:downstream}}

Having developed an efficient $O(n^3)$ dynamic program to compute $\qpsi$, we further leverage the perspectives of \cite{Fornace2020-unified} and \cite{Fornace2025-new} to develop related algorithms for other statistical quantities in \cref{s:quantities}, each operating in the same or lower computational complexity. 

\subsection{Sequence sampling\label{s:sampling}}

The simple intuition behind sampling is that a partition function computed by dynamic programming may be efficiently \emph{backtracked} in order to yield a direct sampler over all of the summed-over degrees of freedom. 
For a fixed RNA sequence, one often uses sampling to compute empirical expectations of arbitrary observables in the Boltzmann ensemble.
In our context, we will use it to perform sequence design with only modest modifications.

Indeed, we need only extend the sampling to yield not only a secondary structure $s$ but also 
the RNA sequence $\phi$ itself.
We describe our detailed algorithms and give pseudocode in \cref{si:pseudocode}.


\subsection{Equilibrium base pair probabilities}
Substitution of $\DesignFold$ for $\Fold$ in \cite{Fornace2025-new}  yields the equilibrium base pair probability matrix, averaged over all RNA sequences, $\Ppsi$, cf. \cref{eq:pair-probability}.
This matrix offers insight as to the main structural features dominating a given codon optimization problem.
In summary of \cref{si:reverse}, we find that:
\begin{eqn}\label{eq:pair-formula}
    \Ppsi_{i-1,j} &= \TrIS[Q_{i,j} R_{i,j}^\t] \exp(\qpsi), \\ &\text{ where }     R_{i,j} = \begin{cases}
        Q_{i+\Nphi,j} & i \geq j \\ 
        Q_{j,i+\Nphi} & \text{otherwise}
    \end{cases},
\end{eqn}
where \cref{eq:pair-formula} is analogous to the respective formula developed in \cite{Fornace2025-new} for a fixed sequence $\phi$.
The matrices $Q$ and $R$ may be calculated with $O(n^3)$ cost before the $O(n^2)$ evaluations of \cref{eq:pair-formula}.


\begin{figure*} \input{floats/histograms.tex} \end{figure*}

\subsection{Marginal computation\label{s:marginal}}

We next show how problems of \emph{marginal} computation can be \emph{exactly} solved, altogether incurring a $O(n^2)$ cost after the $O(n^3)$ cost of free energy computation. 
Beyond statistical interpretation, marginal probabilities enable the design of per-codon (or per-base) bonuses to fit any collection of target frequencies, a possibility we explore in \cref{si:iterative-reweighting}.

To ground our approach, consider (1) that contributions from the last base $\phi_{n-1}$ appear only in the final recursion \cref{eq:complex-logq} and (2) that any base may be made to appear ``last'' by rotating our frame of reference.
Consequently, we compute the partition functions of all ranges of up to length $n$, \emph{including} ranges which wrap around the periodic boundary condition of $0 \equiv n$ (cf. \cref{si:reverse}), incurring $O(n^3)$ cost.
Then the marginal base probability at position $i$ is just:
\begin{eqn} \label{eq:base-marginal}
    \pbase_i(b) = e^{\qpsi} \hspace{-1em} \sum_{k \in \Range{i}{i+\Nphi}} \TrIS[Q_{i,k} \, (\DesignFold |_{\phi_i = b})_{k,i+\Nphi}^\t(Q_{k+1,i+\Nphi-1})],
\end{eqn}
where $\DesignFold |_{\phi_i = b}$ denotes the $\DesignFold$ function modified to only allow indices compatible with $\phi_i = b$.

Next, to calculate the marginal codon probability at amino acid position $i$, we observe that a single codon's contributions may be isolated by manipulating tensor $T_{3i+1}$, i.e. that of the middle base in that codon. 
Thus, we compute:
\begin{eqn} \label{eq:codon-marginal}
    \pcodon_i \hspace{-0.15em} (c) \hspace{-0.15em} = \hspace{-0.2em} e^{\qpsi} \hspace{-1.2em} \sum_{k \in \Range{j}{j+\Nphi}} \hspace{-1em} \TrIS[Q_{j,k} (\DesignFold |_{ \sigma_j = c})_{k,j+\Nphi}^\t (Q_{k+1,j+\Nphi-1})] ,
\end{eqn}
where $j=3i+1$ and $\DesignFold |_{ \sigma_j = c}$ denotes the $\DesignFold$ function modified to only allow indices compatible with $\sigma_j = c$.
Evaluation of \cref{eq:base-marginal} or \cref{eq:codon-marginal} has just $O(n)$ cost for any given $i$.


\subsection{Minimization\label{s:minimization}}

In this section, we modify the partition function computation of \cref{s:free-energy} to compute a minimum free energy (MFE) triplet $(\phi, s, \zeta)$ solving  \cref{eq:min-problem}.

\begin{table}
    \footnotesize
    \centering
\setlength{\tabcolsep}{1.4pt}
\begin{tabular}{c|rrrrrrrrr}
\toprule
Sys. & $\Abs{\psi}$ & $\TotalDim$ & Prefactor & $T_g$ & $T_\mathrm{sample}^{2500}$ & $S_\mathrm{sample}^{2500}$ & $T_\mathrm{rev}$ & $T_{\min}$ & $T_\mathrm{argmin}$ \\
\midrule
1 & 179 & 1025 & 7.0 (1.6) & 0.33 & 0.36 & 24 & 0.35 & 1.0 & 0.58 \\
2 & 643 & 3374 & 5.3 (2.4) & 3.2 & 1.7 & 21 & 3.0 & 63 & 0.78 \\
3 & 1272 & 6712 & 5.4 (2.6) & 14 & 3.7 & 19 & 14 & 335 & 1.4 \\
4 & 2409 & 13664 & 6.8 (3.2) & 74 & 7.7 & 34 & 86 & 2567 & 4.8 \\
\bottomrule
\end{tabular}
\caption{\label{tab:runtime-table}
    Runtime details for example systems in \cref{s:studies}.
{
    \normalfont
    Number of amino acids, dimension of resulting $Q$ matrix, theoretical slowdown of design over analysis (assuming $O(n^3)$ complexity, with the measured slowdown parenthesized), time to compute $\qpsi$, additional time to sample 2500 sequences, speedup factor resulting from simultaneous sampling (\cref{si:sample-alg}), time to compute reverse matrix $R$, time to compute $\qmin$, and time to compute $\phi_{\min}$.
    $T_q$ and $T_{\mathrm{reverse}}$ used 2 GPUs, $T_{\min}$ used 64 CPUs, with other timings in serial.
    All timings are wall clock times in seconds.
}
}
\vspace{-2em}
\end{table}

In short, we invoke the evaluation algebra approach of \cite{Fornace2020-unified}, isolating the structure of the recursions from the precise mathematical operations being invoked within this structure.
To perform minimization, we then swap all operations in the classical semiring (\textsc{SumProduct}) of partition function computation with their equivalents in the tropical semiring (\textsc{MinSum}) (e.g., \cite{Pin1998-tropical,Liu2021-tropical}). 
This allows us to directly access the zero-temperature limit of our Boltzmann ensemble over $(\phi,s, \zeta)$.



\section{Computational studies and benchmarks \label{s:validation}}

In this section, we demonstrate and examine the performance of our algorithms on multiple sequence design problems of real-world interest.
For these problems, we demonstrate that our sampling methodology produces ensembles of sequences that are systematically more stable than those produced by uniform sampling or more na\"ive methods. 
We also demonstrate that our surrogate minimization approach yields sequences of extremely low free energies.
We discuss which outputs are most useful for practical deployment.
Finally, we report the empirical cost of our methods.

\begin{figure*} \input{./floats/performance.tex} \end{figure*}

\subsection{Example proteins and free energy models\label{s:studies}}

We demonstrate our algorithms for codon optimizations defined by the following proteins:
\begin{enumerate}[itemsep=0pt] 
    \item \textit{Cancer/testis antigen 1} (NY-ESO-1 \cite{uniprot-P78358}; UniProt: \\P78358, 179 aa, 537 nt), an immunotherapy target expressed in a variety of tumors \cite{Thomas2018}
    \item \textit{Protein ecdysoneless homolog} (ECD\_HUMAN \cite{uniprot-O95905}; \\UniProt: O95905, 643 aa, 1929 nt), a cell cycle regulator involved in tumor suppression \cite{Mir2016}
    \item \textit{Spike glycoprotein} (SPIKE\_SARS2 \cite{uniprot-P0DTC2}; UniProt: \\P0DTC2, 1272 aa, 3816 nt),  the most recognizable part of the SARS-CoV-2 virus and a past vaccine target \cite{MartinezFlores2021}
    \item \textit{Lipomycin polyketide synthase} (LipPKS SY172 \cite{jbei-pAN001-SY172}; 2409 aa, 7227 nt), a catalytic target of current metabolic engineering, engineering applications \cite{Schmidt2023-maximizing}
\end{enumerate}
We compare our computed RNA sequences against wild-type versions. 
\cref{fig:studies} shows the free energies of the ensembles of sequences designed by the various approaches.
For simplicity, in the main text we exclude results incorporating CAI and CPB bonusing, which are deferred to \cref{si:studies}.

For our main results, we use the $(\blocks=2,\blocksize=3,\Np=1)$ RNA free energy model parameters of \cite{Fornace2025-new} (see \cref{si:study-details} for full parameter details).
For comparison, we also consider the $(\blocks=1,\blocksize=1,\Np=1)$ RNA parameters, which furnish a highly simplified free energy model equivalent to a summation of scalar free energies for each base pair.
Sequences designed using the simplified model are substantially less stable when evaluated with the more accurate free energy model.
Therefore, we conclude that simplified free energy models do not suffice for optimizing the real-world stability of an RNA sequence coding for a given protein.
Likewise, sequences sampled uniformly without regard to structural stability are fairly unstable in general.
In summary, our methods offer consistently more stable results than simpler heuristics.



\subsection{Performance\label{s:performance}}

Overall, in our approach, we are able to leverage the basic skeleton and algorithms for tensor-based sequence analysis for the problem of stability-based sequence design.
As such, we can take advantage of the same efficient computational techniques developed in \cite{Fornace2025-new}.
In summary, we parallelize computations of the folding operation (\cref{eq:dp-short}, first line) over multiple CPU cores and perform the dominant computations of (\cref{eq:dp-short}, second line) on GPUs.
In this work, we exhibit multi-GPU acceleration for the first time.
\Cref{tab:runtime-table} summarizes the expenses of our example studies.
We also benchmarked our algorithms on a set of randomly generated amino acid sequences (see \cref{si:benchmark} for details).
For these, \cref{fig:performance} shows the speedups from GPU acceleration and the comparison in cost to sequence analysis for a range of randomly chosen amino acid sequences.

The free energy algorithm runs in $O(\TotalDim^3)$ time with $O(\TotalDim^2)$ space, where $\TotalDim \Eq \sum_{i \in \UpTo{n}} \DimTT_i$, a factor easily calculated by inspection of the sequence tensor train.
Across our benchmark studies, $\TotalDim / n$ is between 1.7 and 1.9.
We expect this factor to be roughly constant for any problems using the single codon tensor train with the standard codon table.
In practice, we observe that (1) codon optimization takes only \LessPrefactor{4} as long as sequence analysis and (2) GPU acceleration yields up to $\sim$100$\times$ speedups over serial computation. 
See \cref{si:parallelization} for performance and parallelization details and \cref{si:alternative-methods-comparisons} for alternative method comparisons.


\section{Conclusion and future directions \label{s:discussion}}

In this work, we have proposed a principled approach to codon optimization based on dynamic programming, tensor modeling, and statistical sampling.
Because our methodology can apply to relatively arbitrary bonusing functions locally built in terms of sequence or secondary structure, we expect it to be highly compatible with data-driven approaches to codon optimization.
What our methodology offers, from this point of view, is a way to translate a diverse and powerful tensor-based model of secondary structure and sequence ensembles into exactly and efficiently computed sequence designs.

The precise ingredients and parametrizations of our algorithmic components may be changed without changing the essential logic of our approach.
For instance, we and others are working on a full absorbance-based regression, fine-tuning, and validation of tensor parameters.
The parameters used in this work may be improved in the future without changing the presented algorithms.

In future work, we will explore the optimization of conditional structural objectives, e.g., objectives based on equilibrium base pair probabilities.
While such objectives are necessarily more complicated than sequence stability maximization, we believe the outlined approach to be extensible in this direction.
We envision that a simple iterative approach to sequence design will enable more flexible objective optimization than current local search alternatives.
Thus, we hope that our perspective can lead to a new paradigm for computational sequence design overall.

\subsection*{Code availability and hardware details\label{s:computational-details}}

Prototype algorithms were programmed in Python, while parallelized CPU and CUDA algorithms were programmed in C++17. 
Both codes will be released open-source to the public in the near future.
All computational benchmarks were run on the NERSC Perlmutter cluster (CPU: AMD EPYC 7763, GPU: Nvidia A100) with parallelism up to 64 CPU cores and 2 GPUs (see \cref{si:computational-details} for more details). 

\subsection*{AI usage}
Claude Sonnet 4.6 (Anthropic) was used to proofread this manuscript.

\subsection*{Acknowledgments}
This work was supported in part by the Applied Mathematics Competitive Portfolios program (M.F. and M.L.), the Luis W. Alvarez Fellowship in Computing Sciences (M.F.), and the Laboratory Directed Research and Development Program of Lawrence Berkeley National Laboratory (M.F. and C.W.), all funded by the U.S. Department of Energy's Office of Advanced Scientific Computing Research under Contract No. DE-AC02-05CH11231.
This research used resources of the National Energy Research Scientific Computing Center (NERSC), a Department of Energy User Facility using NERSC award ASCR-ERCAP 0033070. M.L. was also partially supported by a Sloan Research Fellowship.


\bibliography{./references}

\clearpage
\onecolumn
\appendix
\input{si.tex}

\end{document}

%% file: preamble.tex
\usepackage{mathtools}
\usepackage{orcidlink}
\usepackage{amsmath}
\usepackage{amssymb}
\usepackage{booktabs}
\usepackage{enumitem}
\usepackage{float}
\usepackage{placeins}

\usepackage{hyperref}
\hypersetup{
    colorlinks,
    citecolor=blue,
    filecolor=black,
    linkcolor=black,
    urlcolor=blue
}

\usepackage[protrusion=true, expansion=true, stretch=30, shrink=30]{microtype}
\usepackage{algorithm,algpseudocodex,multirow}
\usepackage{comment}
\usepackage{xr}
\usepackage{titlesec}
\usepackage{todonotes}

\usepackage{blindtext}

\newenvironment{eqn}{\begin{equation}\begin{aligned}}{\end{aligned}\end{equation}\ignorespacesafterend}

\newcommand{\lrb}[1]{{\left \{ #1 \right \} }}
\newcommand{\lrp}[1]{{\left ( #1 \right ) }}
\newcommand{\lrs}[1]{{\left [ #1 \right ] }}
\newcommand{\Bullet}[1]{\textbf{#1.}}

\usepackage{cleveref}
\crefname{Section}{Section}{Sections}
\Crefname{Section}{Section}{Sections}
\crefname{subsection}{Section}{Sections}
\Crefname{subsection}{Section}{Sections}
\crefname{subsubsection}{Section}{Sections}
\Crefname{subsubsection}{Section}{Sections}

\newcommand{\Prob}{p}

\DeclarePairedDelimiter\Abs{\lvert}{\rvert}%
\DeclarePairedDelimiter\Norm{\lVert}{\rVert}%

\newcommand{\Ppsi}{P^{\psi}}
\newcommand{\pbase}{p^\phi}
\newcommand{\pcodon}{p^\psi}
\newcommand{\Bias}{\lambda}
\newcommand{\TensorIdx}{\zeta}
\newcommand{\lp}{o}

\newcommand{\DimTT}{\Sigma}
\newcommand{\TotalDim}{\bar{\Sigma}}

\newcommand{\Eq}{\coloneqq}

\newcommand{\Tr}{\mathrm{Tr}}

\newcommand{\Id}{\mathbb{I}}

\newcommand{\Fold}{\ensuremath{\chi}}
\newcommand{\DesignFold}{\ensuremath{\hat{\chi}}}
\newcommand{\Range}[2]{[#1 \mathord{:} #2)}
\newcommand{\UpTo}[1]{[#1]}
\newcommand{\RealRange}[2]{[#1 \mathord{:} #2]}
\newcommand{\Inclusive}[2]{[#1 \mathord{:} #2]}
\newcommand{\Exclusive}[2]{(#1 \mathord{:} #2)}

\newcommand{\Ix}{\mathcal{I}}
\newcommand{\Np}{\xi}
\newcommand{\blocks}{\nu}
\newcommand{\blocksize}{\mu}
\newcommand{\V}{V}
\newcommand{\Pp}{\rho}
\newcommand{\W}{S}
\newcommand{\TrS}{\Tr_{\W}}
\newcommand{\Kron}{\otimes}
\newcommand{\TrIS}{\Tr_{\Id \Kron \W}}

\newcommand{\ov}[1]{#1}
\newcommand{\logq}{g}
\newcommand{\qpsi}{g_\psi}
\newcommand{\qmin}{\logq_\psi^\mathrm{min}}
\newcommand{\qseq}{\logq^{\mathrm{seq}}}
\newcommand{\ologq}{\ov{\logq}}
\newcommand{\Nphi}{n}
\newcommand{\Npsi}{\Abs{\psi}}
\newcommand{\Dim}{\gamma}
\newcommand{\Dims}{\UpTo{\gamma}}

\newcommand{\Dist}{\mathcal{D}}
\newcommand{\SttDist}{\mathcal{U}}

\newcommand{\SeqEnsemble}{\Phi}
\newcommand{\StructureEnsemble}{\Gamma}
\newcommand{\StateEnsemble}{\mathcal{Z}}

\newcommand{\Queue}{\mathcal{M}}

\newcommand{\PartialStrucs}{\vec{s}}

\newcommand{\gap}{g^\mathrm{gap}}

\usepackage{scalerel}
\renewcommand{\t}{{\protect\scaleobj{0.8}{\top}}}
\newcommand{\Mo}{{-1}}
\newcommand{\Mh}{{-1/2}}

\newcommand{\BaseA}{\texttt{A}}
\newcommand{\BaseC}{\texttt{C}}
\newcommand{\BaseG}{\texttt{G}}
\newcommand{\BaseU}{\texttt{U}}

\newcommand{\Vz}{\tilde{V}}
\newcommand{\Bz}{\tilde{B}}

\newcommand{\Codon}{\hat{\sigma}}
\newcommand{\ObservedPrefactor}{4}

%% file: floats/secondary-structures.tex
\centering
\includegraphics[width=\columnwidth]{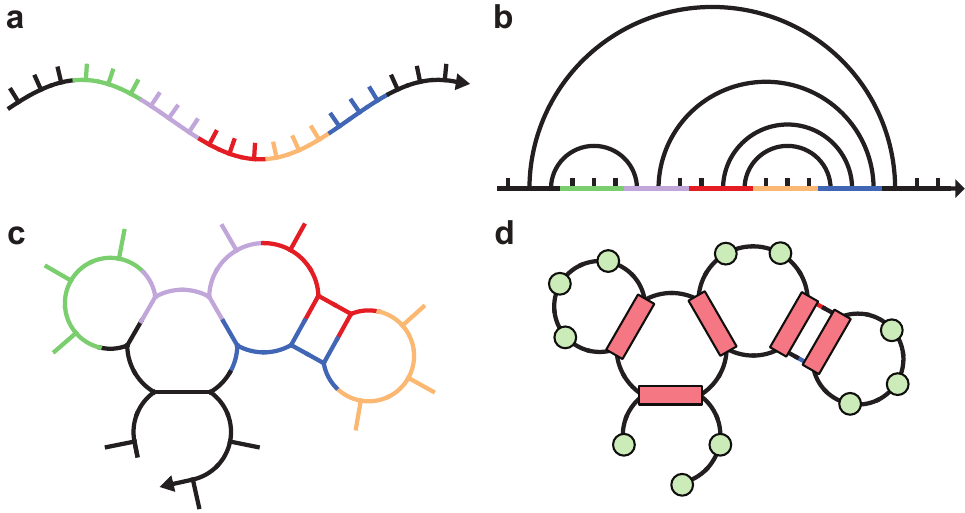}
\caption{\label{fig:model}
       \Bullet{a} An example mRNA strand colored by codon (word of three consecutive bases), with an arrowhead marking the 3$'$ end. 
       The first and last codons (black) are the start and stop codons, while each interior codon codes for a specific amino acid.
       \Bullet{b} Base pair diagram for an example secondary structure with the same sequence as (a), in which the strand is laid out horizontally from 5$'$ to 3$'$ and base pairs appear as arcs between bases.
       Pseudoknots appear as crossing arcs in this type of diagram.
       \Bullet{c} Secondary structure diagram equivalent to (b), formed by contorting (b) such that the backbone curves around clockwise, unpaired bases appear as ticks, and base pairs appear as short line segments. 
       \Bullet{d} Tensor network model for the partition function of the secondary structure shown in (b-c). 
       Each circle indicates a matrix associated with an unpaired base $i$, while each rectangle indicates a four-legged tensor associated with a base pair $i \cdot j$. 
       Each joining segment indicates an index to be summed over in order to compute the secondary structure partition function.
}

%% file: floats/tensor-contraction.tex
\includegraphics[width=\columnwidth]{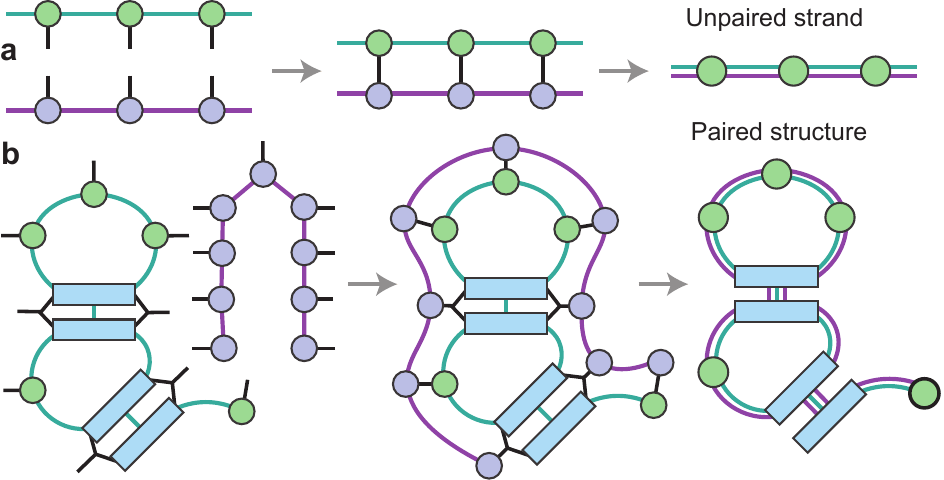} 
   \caption{\label{fig:contraction}
      Combination of structure and sequence tensor diagrams.
      \Bullet{a} Combination of diagrams for an unpaired secondary structure.
      In this case, the structure free energy is simply a matrix product (left, green).
      Combining it with an RNA sequence tensor train (left, purple) yields a compound diagram (middle).
      Then pre-contraction over vertical edges yields another matrix product of expanded dimension (right).
      \Bullet{b} Combination of diagrams for a secondary structure involving loops.
      Here, the structure free energy is a contraction over unpaired bases (green) and two tensors for each base pair (blue).
      The combination of diagrams can be visualized (middle) by the same approach, introducing hyperedges to each side of a given base pair.
      Again, pre-contraction over the connecting edges gives a tensor diagram identical in topology to the initial structure diagram, but with expanded dimension (right). 
   }

%% file: floats/histograms.tex
\centering
\includegraphics[width=0.96\textwidth]{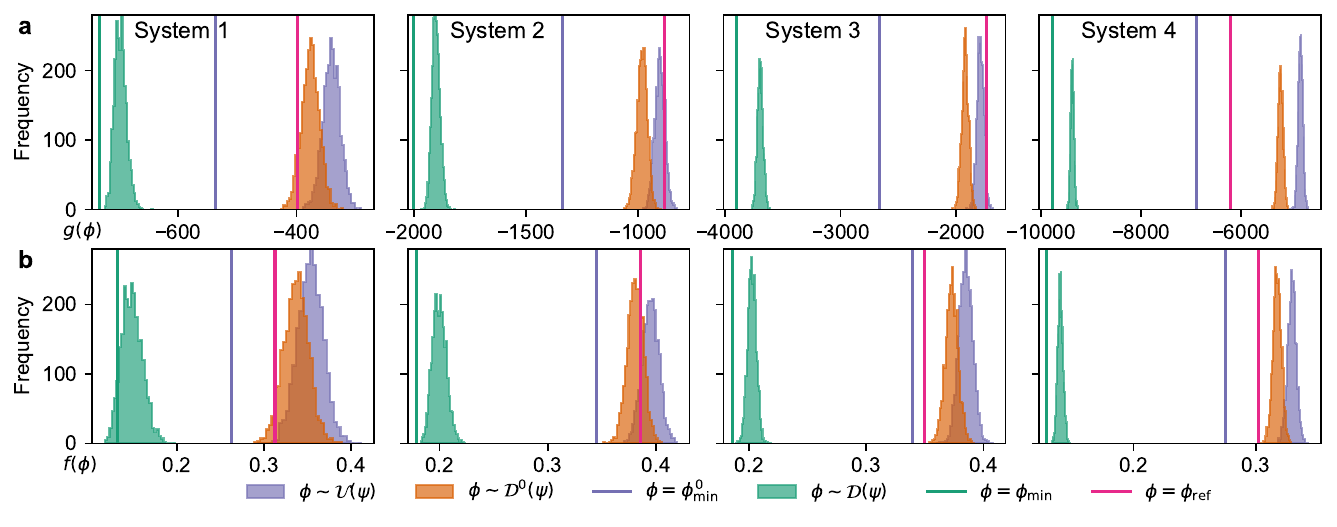}
\vspace{-0.15in}
\caption{\label{fig:studies} Results for example systems listed in \cref{s:studies} (1-4) using different RNA sequence generation methods.
    $\phi \sim \SttDist(\psi)$ was generated by sampling 2500 codon sequences independently (uniformly subject to codon constraints, with no consideration of secondary structure).
    $\phi \sim \Dist(\psi)$ was generated by sampling 2500 codon sequences proportional to $e^{-\qpsi(\phi)}$, while
    $\phi_{\mathrm{min}}$ is the optimum yielded by solving \cref{eq:min-problem}.
    $\Dist^0(\psi)$ and $\phi_{\mathrm{min}}^0$ reflect the same quantities but using a simplistic model based on additive base pair free energy bonuses.
    $\phi_{\mathrm{ref}}$ is a reference wild-type sequence.
\textbf{a}: complex free energy $g(\phi)$ (unitless; multiply by 0.62 kcal/mol for results at 37\textdegree C).
\textbf{b}: equilibrium fraction of unpaired bases $f(\phi)$.
No soft constraints were applied; see \cref{si:studies} for additional studies incorporating CAI and CPB bonuses.
}

%% file: floats/performance.tex
\centering
\includegraphics[width=\linewidth]{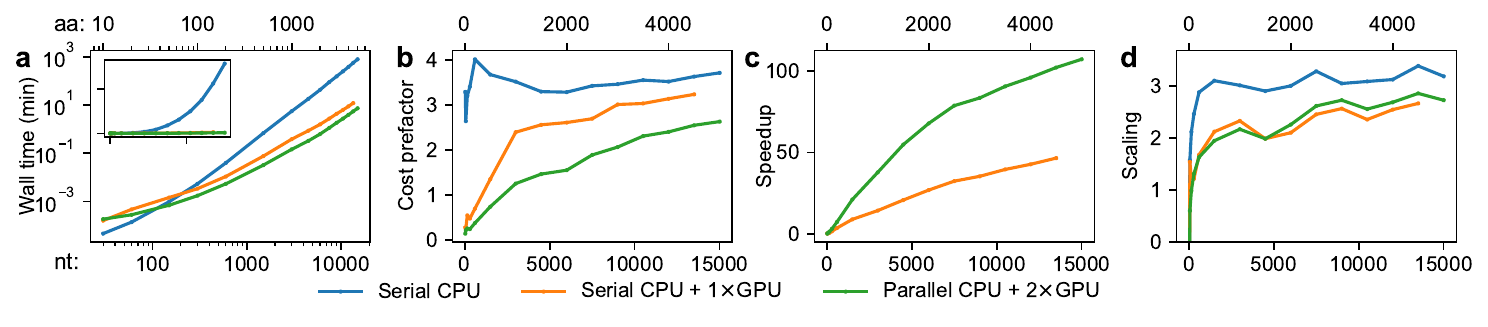}
\vspace{-2em}
\caption{
Algorithm performance and observed complexities in computing $\qpsi$ as a function of sequence length (in nucleotides (nt) on bottom and in amino acids (aa) on top), illustrating the effects of GPU acceleration and parallelization.
\Bullet{a} Wall clock time to compute $\qpsi$.
Inset: same, with linear axes.
\Bullet{b} Cost of computing $\qpsi$ divided by the cost of computing $\qpsi(\phi)$ for an equal length sequence, showing that design is within a cost factor of $\approx 4$ of analysis for this problem.
\Bullet{c} Speedup of parallelized and GPU algorithms over serial CPU.
\Bullet{d} Empirical computational complexity by approximating the slope of $\log t$ vs $\log n$ via backwards finite difference. 
While all algorithms have a theoretical complexity of $O(n^3)$, GPU acceleration ameliorates the empirical complexity to $\approx O(n^2)$ or lower for many problem sizes.
See \cref{s:performance} and \cref{si:benchmark} for more details. 
\label{fig:performance}}

%% file: si.tex



\renewcommand*{\thesection}{S\arabic{section}}
\renewcommand*{\thesubsection}{S\arabic{section}\alph{subsection}}

\newcommand{\Freq}{f}


\clearpage
\section{Glossary of symbols \label{si:notation}}


We provide a glossary of notation below.

\begin{table}[H]
    \small
\centering
\renewcommand{\arraystretch}{1.2}
\begin{tabular}{c|p{6in}}
\toprule
    Symbol & Definition \\
\midrule
    $\psi$ & the amino acid sequence (of length $\Npsi$) being coded for (a fixed input of our algorithms) \\
    $\phi$ & an RNA sequence of length $\Nphi$ (so that $n = \Abs{\phi}$), generally satisfying $\Nphi = 3 \Npsi$ \\
    $s$ & an RNA secondary structure (representable as a list of base pairs) \\
    $i \cdot j$ & for RNA sequence indices $i$ and $j$, the condition that $i$ and $j$ are paired, if $i \neq j$, or that $i$ is unpaired, if $i = j$ \\
    $\zeta$ & a structural configuration compatible with a given RNA secondary structure, equivalent to a full set of tensor contraction indices in the respective tensor diagram.
    $\zeta_x$ refers to a single contraction index in this context (for a specified $x$). \\
    $\SeqEnsemble(\psi)$ & the set of all RNA sequences which code for amino acid sequence   $\psi$ \\
    $\StructureEnsemble(\phi)$ & the set of all secondary structures compatible with sequence  $\phi$ \\
    $\StateEnsemble(s)$ & the set of all configurations compatible with secondary structure $s$ \\
\midrule
    $\Sigma_i$ & the dimension of the edge between tensors $i-1$ and $i$ in the RNA sequence tensor train \\
    $\sigma_i$ & is an index in the RNA sequence tensor train, specifically for the edge between tensors $i-1$ and $i$  \\
    $\Codon_k$ & the $k$th codon represented by a given RNA sequence. The codon sequence $\Codon \Eq (\Codon_0, \dots, \Codon_{\Npsi-1})$ is in bijection with the RNA sequence $\phi$ and is generally in bijection with any feasible sequence of tensor train indices $\sigma \Eq (\sigma_0, \dots, \sigma_{n-1})$. \\
\midrule
    $\logq(\phi, s, \zeta)$ & the unitless free energy of sequence $\phi$ in structure $s$ and configuration  $\zeta$ \\
    $\logq(\phi, s)$ & the unitless free energy of sequence $\phi$ in structure $s$ \\
    $\logq(\phi)$ & the unitless free energy summing over all structures $s$ compatible with $\phi$ \\
    $\Dim$ & the essential dimensionality of the tensor-based free energy model for a fixed RNA sequence, equal to the dimension of each edge of any two tensors in the structure tensor diagram (except for the edges connecting two $B$ tensors representing two sides of the same base pair) \\
    $\Np$ & the base pair correlation rank, equal to the dimension of the edge between any two $B$ tensors on either side of the same base pair \\
\midrule
    $\qpsi(\phi, s, \zeta)$ & same as $\logq(\phi, s, \zeta)$ but $\infty$ for any $\phi \notin \Phi(\psi)$ \\
    $\qpsi(\phi, s)$ & same as $\logq(\phi, s)$ but $\infty$ for any $\phi \notin \Phi(\psi)$ \\
    $\qpsi(\phi)$ & same as $\logq(\phi)$ but $\infty$ for any $\phi \notin \Phi(\psi)$ \\
    $\qpsi$ & the free energy summing over all sequences $\phi$ compatible with $\psi$ and  structures $s$ compatible with $\phi$ \\
    $\qpsi^\mathrm{seq}(\phi)$ & the free energy yielded by the sequence tensor train (which incorporates all sequence-based hard and soft constraints) \\
    $\Dist(\psi)$ & the probability density distribution over states $(\phi,s,\zeta)$ according to free energy $\qpsi(\phi, s, \zeta)$. We implicitly  marginalize over unnecessary variables, i.e. $\phi \sim \Dist(\psi)$ means sampling a tuple $(\phi,s,\zeta)$ and returning just $\phi$. \\
    $\SttDist(\psi)$ & the probability density distribution over sequences with respect only to the free energy $\qpsi^\mathrm{seq}(\phi)$ \\
    $\Bias_\psi(\phi)$ & the soft constraint free energy derived from sequence-based objectives (e.g., CAI or CPB) \\
\midrule
    $\Ppsi_{i,j}$ & marginal probability of a base pair/unpaired base $i \cdot j$ over distribution $\Dist(\psi)$ \\
    $\pbase_i(b)$ & marginal probability of base $i$ being equal to $b$ \\
    $\pcodon_i(c)$ & marginal probability of codon $c$ being sampled for amino acid $i$ \\
\midrule
    $Q_{i,j}$ & the partition function matrix corresponding to the range of nucleotides $\Range{i}{j}$ if $i \leq j$ or the partition function matrix corresponding to the range of nucleotides $\Range{j}{i}$ constrained such that $(i-1) \cdot j$ if $j < i$ \\
    $R_{i,j}$ & the partition function corresponding to $Q_{i+\Nphi,j}$, if $i \leq j$, or $Q_{j,i+\Nphi}$, if $i < j$ (recalling our use of cyclic indexing, so that $\Range{j}{i}$ for $i < j$ wraps around from $j$ to $i$ in a 5$'$ to 3$'$ direction) \\
    $\Fold$ & the folding function used in dynamic programs for a fixed RNA sequence \\
    $\DesignFold$ & the folding function used in dynamic programs for a variable RNA sequence \\
\midrule
    $V_i$ & tensor corresponding to an unpaired base $i$, part of the tensor-based free energy model \\
    $B_{i,j}$ & tensor corresponding to one side of a base pair $i \cdot j$, part of the tensor-based free energy model \\
    $S$ & global trace matrix, part of the tensor-based free energy model \\
    $\TrS$ & weighted trace according to $S$, so that $\TrS[X] = \Tr[S X]$ for any square matrix $X$ \\
    $\Tr_{\Id \otimes S}$ & the same, but replacing $S$ with $\Id \otimes S$ to sum structural degrees ($\zeta$) of freedom by $S$ and sequence ones ($\sigma$) by $\Id$, cf. \cref{fig:method} \\
    $T_i$ & tensor at position $i$ in the RNA sequence tensor train \\
    $\tau$ & an operator specifying a single entry in one of the sequence tensor train tensors \\
\bottomrule
\end{tabular}
\end{table}

\clearpage
\section{Methodological background}

In this section, we summarize the background and motivations for our current work in more detail.
In general, our current work builds heavily on a number of existing paradigms and shows that they generalize very nicely to our codon optimization problem.

\subsection{Secondary structure free energy models}

The thermodynamics of nucleic acids may be modeled relatively effectively in terms of secondary structure alone.
As a result, different free energy models for secondary structures have been developed over the last five decades (e.g., \cite{Tinoco1971-estimation,Serra1995-predicting,SantaLucia1998-unified,Mathews1999-expanded,Turner2010-nndb}).
In most cases, for algorithmic reasons, researchers preclude the consideration of pseudoknotted secondary structures, which contain crossing base pairs.
This approximation is known to be relatively accurate for many classes of smaller or designed nucleic acid systems.
On the other hand, certain large nucleic acid complexes may be significantly pseudoknotted in reality.
In the current work, we also exclude pseudoknotted structures from consideration, though we anticipate future extensions to approximately incorporate pseudoknots.

\emph{Nearest-neighbor secondary structure models} comprise the predominant class of free energy models used in practice.
In these models, an unpseudoknotted secondary structure is decomposed into a set of loops, and the free energy of the secondary structure is approximated as a sum of free energies for each loop.
The free energy of each loop, in turn, is formulated in a complicated way, depending on the class of loop (hairpin, exterior, etc.) and involving various lookup tables, functional forms, and experimentally estimated parameters.
\cite{Fornace2020-unified} gives a thorough description of the nearest-neighbor models for RNA and DNA as implemented in NUPACK 4 (and points to additional relevant literature).
While nearest-neighbor models have been developed and used for decades, their level of complication hinders future developments (including reparametrizations, extensions to new materials, and GPU acceleration).
In particular, we believe that nearest-neighbor models are fundamentally incompatible with the algorithms presented in this paper.





\subsection{Tensor-based free energy models}

We recently proposed and benchmarked a class of \emph{tensor-based free energy models} for nucleic acid thermodynamics in \cite{Fornace2025-new}.
This class of model is broadly intended to replace nearest-neighbor models, i.e., to provide an accurate and efficient model of secondary structure free energies for a given nucleic acid sequence.
In contrast to nearest-neighbor models, the ingredients of our tensor-based models are relatively simple:
\begin{enumerate}
    \item For each unpaired base $i$, a matrix $\Vz_i$
    \item For each base pair $i \cdot j$, a pair of tensors $\Bz_{i,j}$ and $\Bz_{j,i}$ modeling each side of the base pair
    \item A matrix $Z$ modeling each strand break in secondary structure, constructed as $Z = l r^\t$ for global vectors $l$ and $r$
    \item A matrix $S$ used to yield a weighted trace $\TrS[X] \Eq \Tr[S X]$ while maintaining the invariance of the resulting model under cyclic permutation
\end{enumerate}
To elide the use of $Z$ in our presentation, we define $V$ and $B$ tensors, which fold $Z$ into $\Vz$ and $\Bz$ (respectively) as appropriate.

From these ingredients, the free energy of a given secondary structure and nucleic acid sequence is formulated in terms of the tensor diagram matching that structure (\cref{fig:model}).
(\cref{fig:model}d simplifies our presentation just by pre-contracting each pair of $B$ tensors before depiction.)
Generally, the free energy in this model is:
\begin{eqn}
    \logq(\phi, s) = -\log \sum_{\TensorIdx_1} \cdots \sum_{\TensorIdx_m} X_1^{(\TensorIdx)} \cdots X_n^{(\TensorIdx)},
    \label{eq:tensor-free-energy}
\end{eqn}
where $(X_1, \dots, X_n)$ are the $n$ tensors in the diagram associated to secondary structure $s$ (e.g., \cref{fig:structures}b), $(\zeta_1, \dots, \zeta_m)$ is the sequence of configuration indices, one for each edge in the diagram, and $X_i^{(\zeta)}$ denotes the scalar element obtained by addressing $X_i$ with its respective subset of the configuration indices $\zeta$.

More specifically, we can directly write the free energy in terms of $V$ and $B$ tensors as:
\begin{eqn} \label{eq:general-model}
    \logq(\phi, s) 
    &= -\log \sum_{\Pp_1 \in \UpTo{\Np}} \hspace{-0.1em} \cdots \hspace{-0.1em} \sum_{\Pp_{m} \in \UpTo{\Np}}
    \prod_{\lp \in \mathrm{loops}(s)} \hspace{-0.5em} \TrS \hspace{-0.2em} \left[ \prod_{i=1}^{\Abs{\sigma}} 
    \begin{cases}
        \V_{\lp_i} & \text{base } \lp_i \text{ is unpaired} \\
        B_{\lp_i, \lp_{i+1  }}^{\Pp(\lp_i)} & \text{base } \lp_i \text{ is paired to base } \lp_{i+1}   \\
        \Id & \text{otherwise}
    \end{cases}  \right], 
\end{eqn}
yielding a single equation for the secondary structure free energy.
Here, the outer summations over $\rho_1,\dots,\rho_m$ allows for correlations between loop structural ensembles, making the tensor-based description more flexible than nearest-neighbor models at an essential level.
For a fixed set of $\rho$ indices, meanwhile, the partition function of a loop is formulated as the trace of a product of each matrix along the backbone of the loop.
This trace formulation\footnote{
    In fact we use the weighted trace $\TrS[X] \Eq \Tr[\W X]$, where $S$ is a global matrix defined as part of the free energy model. 
    We use this notation as the model is specifically constructed to enforce cyclic symmetry of $\TrS$, implying that $\TrS[ABC]=\TrS[BCA]=\TrS[CBA]$ just as in the normal trace identity that $\Tr[ABC]=\Tr[BCA]=\Tr[CBA]$.
    This is key to the rotational invariance of this type of model.
} is essential to ensuring that the complete free energy model is invariant with respect to rotations of the complex.
For an unpaired strand, this takes on the simple form $\ologq(\phi, s) = -\log \sum \TrS[V_0 \cdots V_{n-1}]$ and is, in this case, exactly analogous to transfer matrix formulations appearing elsewhere (e.g., for protein conformations \cite{Kloczkowski2004} and the classical 1D Ising model \cite{kardar2007statistical}).

In \cite{Fornace2025-new}, we have demonstrated that this class of model can (1) accurately reproduce existing nearest-neighbor models and (2) accurately reproduce literature experimental melt data.
In doing so, we demonstrated that these models could satisfactorily model both local energetic effects (including base pairing and $\pi$-$\pi$ stacking), nonlocal entropy effects, and moreover could be simply constrained to avoid any disconnected complexes (in the case of multistranded complexes).
Furthermore, we demonstrated that these models enabled all of the same sorts of dynamic programming algorithms as was possible with nearest-neighbor models.
Indeed, these algorithms were significantly simplified and improved for our models, and we were able to cast many computations in terms of triangular and rectangular matrix multiplications.
This enabled a high degree of efficiency, as we could leverage accelerated BLAS routines for these operations.
In fact, we demonstrated that GPU acceleration of these routines using CuBLAS provided large speedups of up to 175$\times$ for sequence analysis.


\subsection{Dynamic programming algorithms for fixed RNA sequences}

Exploiting the relatively simple nature of secondary structure models and the prohibition of pseudoknots, numerous dynamic programming algorithms have been developed over the years (e.g., \cite{Dirks2007-thermodynamic,Ding2003-statistical,Zuker1981-optimal,McCaskill1990-equilibrium,Fornace2020-unified}) to compute various thermodynamic observables:
\begin{itemize}\setlength\itemsep{0em}
    \item The \emph{complex free energy} $\logq(\phi)$, the negative logarithm of the partition function summing the Boltzmann factors for all possible (unpseudoknotted) secondary structures for a given sequence $\phi$
    \item The state \emph{minimum free energy} or, more precisely, the minimum of $\logq(\phi, s, \zeta)$ over any state $s, \zeta$ compatible with fixed input sequence $\phi$.
    \item The \emph{equilibrium base pair probabilities}, represented as a symmetric (and doubly stochastic) $(n,n)$ matrix in which each $(i,j)$ element is the equilibrium probability of base pair or unpaired base $i \cdot j$ over the Boltzmann ensemble of possible secondary structures.
    \item A list of \emph{Boltzmann sampled structures}, which are i.i.d. secondary structures drawn with probabilities proportional to Boltzmann factor $e^{-\logq(\phi,s)}$.
    \item A set of \emph{suboptimal structures}, which comprises all secondary structures $s$ for which $\logq(\phi, s, \zeta) \leq \logq_\mathrm{MFE}(\phi) + \gap$.
    Choosing $\gap = 0$ returns all structures corresponding to a state with a free energy equal to $\logq_\mathrm{MFE}(\phi)$ (often just a single structure).
\end{itemize}
Remarkably, these observables can be computed \emph{exactly} in $O(n^3)$ time with respect to this approximate free energy model and reduced structural ensemble, even though the number of possible unpseudoknotted secondary structures scales like $e^{O(n)}$.

In \cite{Fornace2020-unified}, we developed a unified scheme for dynamic programming algorithm development and extension.
In our scheme, we conceived of each dynamic program as combining three programmatic elements:
    (1) a set of recursions defining subproblem relations for the given structural ensemble and free energy model, 
    (2) an evaluation algebra, translating the physical quantity of interest into an abstract mathematical semiring, 
    and (3) an operation order defining an efficient outer iteration over subproblems.
To a large degree, our modular approach was motivated by a desire to isolate the high complication of nearest-neighbor models.
However, in \cite{Fornace2025-new} we demonstrated that this same scheme could be readily extended to tensor-based free energy models, exposing huge opportunities for simplification and acceleration.


\subsubsection{Recursion definition}
In our tensor-based approach, we arrived at a \emph{single} recursion replacing the large set of prior recursions associated with nearest-neighbor models.
While \cite{Fornace2025-new} describe this recursion in full, we give a quick motivation here.
As an elementary step, consider a simple recursion that counts the number of noncrossing secondary structures for a sequence in which any two bases can pair:
\begin{eqn}
    Q^\mathrm{scalar}_{i,j+1} \Eq Q^\mathrm{scalar}_{i,j} + \sum_{k \in \Range{i}{j}} Q^\mathrm{scalar}_{i,k} Q^\mathrm{scalar}_{k+1,j},
    \label{eq:scalar-pf}
\end{eqn}
where $Q^\mathrm{scalar}_{i,j}$ is the number of structures formable by subsequence $\Range{i}{j}$.
Here, the first right-hand term incorporates an additional 3$'$ unpaired base $j$, while the latter summation sums over all base pairs which can be closed by base $j$ (as in \cref{fig:structures}a).
In fact, in the end, we derive a very similar recursion (for any $i \leq j$) for our tensor-based formulation in \cite{Fornace2025-new}:
\begin{eqn}
    Q_{i,j+1} = 
        Q_{i,j} \V_{j} +  \sum_{k \in \Range{i}{j}} \sum_{\Pp \in \UpTo{\Np}} Q_{i,k} B_{k,j}^\Pp \TrS[B_{j,k}^\Pp Q_{k+1,j}],
    \label{eq:complicated-pf}
\end{eqn}
with the main modifications coming from (1) replacing scalar multiplications with matrix multiplications, (2) incorporating our free energy parameter tensors $V$ and $B$, and (3)  using a trace-based summation over each possible base pair.
As a final simplification, we extended our recursion to the case of $j<i$ and defined the folding function $\Fold$, yielding (as in \cref{eq:dp-short,eq:fold}):
\begin{eqn} \label{eq:dp-si}
    Q_{i, j} = \begin{cases}
        \sum_{k \in \Range{i}{j}} Q_{i,k} Q^\t_{j,k} & i < j \\
        \Fold_{j,i}(Q_{j+1,i-1}) & \text{otherwise}
    \end{cases}  \quad \text{where} \quad
    \Fold_{i,j+1}(X) \Eq \begin{cases}
        \Id & i = j+1 \\
        \V_{i}^\t & i = j \\
        \sum_{\Pp \in \UpTo{\Np}} B_{i,j}^{\Pp,\t} \TrS[B_{j,i}^{\Pp} X] & \text{otherwise}
    \end{cases}.
\end{eqn} 
In keeping with our previous work \cite{Fornace2025-new}, we define $B_{i,j} = 0$ for any indices $i,j$ for which both (1) $\Abs{j-i} < 4$ and (2) there is no strand break between bases $i$ and $j$.
This prevents hairpin loops with less than three unpaired nucleotides.

At dynamic program completion, we expand $\TrS[Q_{0,n}]$ (where $n \Eq \Abs{\phi}$) to yield 
\begin{eqn} 
    \logq(\phi) \Eq -\log \sum_{k \in \UpTo{n}} \TrS[Q_{0,k} \, \Fold_{k,\Nphi}^\t(Q_{k+1,\Nphi-1})].
    \label{eq:complex-logq-si}
\end{eqn}
Equations \cref{eq:dp-si,eq:complex-logq-si} comprise our entire approach to complex free energy computation.

\subsubsection{Generalization via evaluation algebras}
\cref{eq:complicated-pf} represented our main recursion to calculate the free energy of a sequence $\logq(\phi)$.
Next, in \cite{Fornace2020-unified}, we developed a concrete mathematical concept of an \emph{evaluation algebra}, which translates a desired physical observable into an abstract set of mathematical operations.
These operations are summarizable as an algebraic semiring.
We used this concept to extend the recursion \cref{eq:dp-si} to the calculation of all other physical quantities of interest.
In essence, in this formalism, addition and multiplication operations are abstracted and substituted based on the quantity being computed (along with downstream definitions of the additive identity element, the multiplicative identity element, and matrix multiplication).
This concept proved useful both for nearest-neighbor models (in \cite{Fornace2020-unified}) and tensor-based models (in \cite{Fornace2025-new}).

\subsubsection{Efficient operation orders}
Finally, we conceived of an operation order as the ``outer loop'' over recursions, designed in order to minimize the total computational expense of the algorithm.
For free energy computations, our tensor-based operation orders in \cite{Fornace2025-new} allow for high parallelism and GPU acceleration.
For base pair probabilities, an extension to ``complementary'' subsequences yields a similarly accelerated algorithm.
Meanwhile, for Boltzmann sampling and suboptimal structures, the relevant operation orders are essentially \emph{backtracking} algorithms, in which a free energy is decomposed into its individual contributions in a repeated fashion.
We observed that frequently, recursion elements in naive algorithms were backtracked through repeatedly.
Therefore, in \cite{Fornace2020-unified} and \cite{Fornace2025-new} we developed sophisticated priority queue-based operation orders for structure generation.
(We generalize these algorithms to the codon optimization problem in \cref{si:sample-alg,si:subopt}.)

\subsection{Sequence-based codon optimality metrics \label{si:soft-constraint-background}}

The design and impact of different codon optimization methods have been extensively studied in past literature. 
In addition to mRNA structural stability, other common design criteria in the literature include metrics such as the Codon Adaptation Index (CAI), Individual Codon Usage (ICU), Codon Context (CC), Codon Pair Bias (CPB), Relative Codon Bias (RCB), and Relative Codon Pair Bias (RCPB) \cite{demissie2025comparative,Coleman2008-virus,Sen2020}. 
A consistent finding is the importance of incorporating multiple criteria for effective optimization, as highlighted by Demissie et al. \cite{demissie2025comparative}. 
In their comparative analysis, the authors provide a survey of different codon optimization tools applying these criteria. Ward et al. \cite{Ward2025} also compared and evaluated several algorithms co-optimizing RNA structure and codon usage. 

In this work, we assign bonus scores to codons and codon pairs using the Codon Adaptation Index (CAI) and Codon Pair Bias (CPB) metrics, which reflect their preferred usage in the host organism. 
CAI is a common measure of codon preference that shows how closely a gene's codons match the preferred codons of highly expressed genes in the host \cite{Sharp1987}. 
On the other hand, codon pair usage is also strongly biased in many organisms \cite{Gutman1989}, which can have significant impact on gene expression \cite{Hegelmeyer2023} and therefore motivates the use of Codon Pair Bias (CPB) as an additional measure. 
The CPB score was proposed by \cite{Coleman2008-virus} to measure how much the frequently observed codon pairs are used in a codon sequence. 
A high CPB index indicates that the sequence uses more of the preferred codon pairs and less of the underrepresented ones \cite{Sen2020}. 
For exact formulae of these metrics, see \cref{si:soft-constraint-defs}.

From a different angle, motif-based sequence design (e.g., \cite{Condon2012}) bonuses or penalizes the appearances of certain words (small subsequences) of contiguous nucleotides or codons in the produced sequence.
While we do not implement motif-based soft constraints in this work, we note here that this formulation is simply compatible with our approach to sequence design.
However, since the dimensionality of the sequence tensor train will depend on the number and lengths of the input motifs, the cost of our method will have to be optimized and benchmarked carefully for this case.


Of course, we have not implemented all possible sequence-based bonuses, and only a subset of these formulations are compatible with our framework. For instance, the scoring metrics Relative Codon Bias and the distribution-based score using Kullback-Leibler divergence (corresponding to Relative Codon Pair Bias) do not fit naturally into our model. However, the primary motivation for introducing such bonuses is to conserve a desired frequency of codons, which our algorithms achieve directly and globally without requiring additional complexity.

\FloatBarrier
\section{Methodology details \label{si:methods}}

In this section, we give full details for how the fundamentals of dynamic programs over sequences can work in practice.
We begin by detailing the renormalization procedure that we use prior to computation, in order to achieve a roughly uniform sampling prior over sequence.
Next, we explain how we embed various soft and hard sequence constraints into our RNA sequence tensor train.
This lets us formalize the way in which CAI, CPB, or other constraints can be handled in the same way by downstream algorithms.
Specifically, we show an approach to constraint encoding based on a general ``operator list'' format.
Finally, we discuss the computational costs of our algorithms with respect to sequence tensor train dimensionality.





\subsection{Free energy model renormalization}

The tensor-based models we considered were all derived from fitting structural observables given fixed RNA sequences \cite{Fornace2025-new}.
These regressions, involving fixed sequences, therefore have no control on the sampling prior for different sequences.
For example, it is possible to have two free energy models, one with $\logq(\BaseA \BaseA \BaseA \BaseA, \texttt{....}) = 100$ and $\logq(\BaseC \BaseC \BaseC \BaseC, \texttt{....}) = 0$, and one with $\logq(\BaseC \BaseC \BaseC \BaseC, \texttt{....}) = 100$ and $\log(\BaseA \BaseA \BaseA \BaseA, \texttt{....}) = 0$.
Although the probabilities of sampling a sequence of $\BaseA$s or $\BaseC$s will differ between these models, the secondary structure distributions of each model, given a fixed sequence, may be identical.
In essence, special care must be taken in order with our models to ensure an (at least approximately) even playing field for each RNA base.

In order to address this concern, in this work we normalized the $\Vz$ and $\Bz$ parameters of each tensor-based free energy model so that the spectral radii (largest-magnitude eigenvalues) of all $\Vz$ matrices ($\Vz_\BaseA$, $\Vz_\BaseC$, $\Vz_\BaseG$, $\Vz_\BaseU$) were all unity.
(Note that these matrices, being entrywise real and non-negative, possess a largest-magnitude eigenvalue which is real and non-negative, by the Perron-Frobenius theorem.)
This procedure (1) does not affect the probability of any secondary structure for a fixed sequence $p(\phi, s)$
and (2) is a simple way to ensure that the free energies of any unpaired strand are approximately 0 in an average asymptotic sense.
To implement this, we simply calculated the spectral radii $r_x$ for each $\Vz$ matrix $\Vz_x$, and performed the substitutions $\Bz'_{x,y} \gets \Bz_{x,y} r_x^\Mh r_y^\Mh$ and $\Vz'_x \gets \Vz_x r_x^\Mo$ before model use.
The base pair probability profiles achieved by our method are good evidence for the validity of our approach, showing that unpaired probabilities are reduced substantially via our sampling method (e.g., \cref{fig:studies}b).




\subsection{Definitions and implementations of example soft constraints \label{si:soft-constraint-defs}}

In this section, we describe our specific implementations of CAI and CAI/CPB soft constraints.
Each soft constraint is constructed to give a bonus function $\Bias_\psi(\phi)$ which can be calculated simply via an RNA sequence tensor train.
Of course, the precise scalings of these soft constraints are simply tunable. 
Finally, we discuss a simple iterative approach to frequency matching in \cref{si:iterative-reweighting}, which we believe obviates most needs for a user-chosen scaling.
Our tensor train formulation is significantly general and admits any soft constraints that are formulated in terms of products of terms which can be calculated in terms of local sequence information.

\subsubsection{Codon adaptation index (CAI)}

The \textit{fitness value} $f_\mathrm{CAI}$ of a codon $\Codon_k$  in expressing an amino acid $\psi_i$ is the ratio of its observed frequency to the observed frequency of the most frequent codon:
\begin{eqn}
\label{eq:cai-fitness-value}
    f_\mathrm{CAI}(\psi_i,\Codon_k) \Eq \dfrac{\Freq_k^c}{\max_{l \in K_i}\Freq_l^c},
\end{eqn}
where $\Freq_i^c$ is the observed frequency of codon $\Codon_i$ and $K_i$ is the set of codons that can be used to express amino acid $\psi_i$ \cite{Sen2020}.

The Codon Adaptation Index (CAI) of a codon sequence is given by the geometric mean of the fitness values of its codons. The CAI of a codon sequence $\Codon = (\Codon_0,...,\Codon_{\Npsi-1})$ for an amino acid sequence $\psi = (\psi_0,...\psi_{\Npsi-1})$ is given by
\begin{eqn}
    \text{CAI}(\psi,\Codon) \Eq \left(\prod_{i=0}^{\Npsi-1} f_\mathrm{CAI}(\psi_i,\Codon_i) \right)^{1/{\Npsi}}.
\end{eqn}
In order to incorporate the CAI soft constraints into our methodology, we applied the following free energy bonus:
\begin{eqn}
    \Bias^\mathrm{CAI}_\psi(\phi) \Eq -\Npsi \log \mathrm{CAI}(\psi, \Codon \text{ implied by } \phi),
\end{eqn}
ensuring a formulation which is extensive with respect to sequence length.

\subsubsection{Codon pair bias (CPB)}

The codon pair score (CPS) is given by the frequency of a codon pair $(\Codon_k,\Codon_l)$ in encoding amino acid pair $(\psi_k,\psi_l)$ relative to what is expected by chance given the frequencies of each codon. The codon pair is overrepresented or underrepresented if the CPS score is positive or negative, respectively. In the formula shown below, $f^c$ represents the observed frequency of codons or codon pairs, depending on the number of subscripts. Similarly, $f^a$ represents the observed frequency of amino acids or pairs of amino acids.
We write
\begin{eqn}
\label{eq:cps-def}
    \mathrm{CPS}(\psi_i, \Codon_k, \psi_j, \Codon_l) = \log\lrp{\frac{\Freq_{k,l}^c \Freq_i^a \Freq_j^a}{\Freq_{i,j}^a \Freq_k^c \Freq_l^c}} &= \log \lrp{ \frac{f_{k,l}}{f_k f_l} } + \mathrm{const}, 
\end{eqn}
where we use ``$\mathrm{const}$'' to absorb all terms not affecting the relative weights of different RNA sequences $\phi$.

The Codon Pair Bias (CPB) score is proposed by \cite{Coleman2008-virus} to measure how much the frequently observed codon pairs are used in a codon sequence. This index is given by the mean of all the CPS in the sequence:
\begin{eqn}
    \text{CPB}(\psi,\Codon) = \dfrac{\sum_{i=1}^{\Npsi-1}\text{CPS}(\psi_{i-1},\Codon_{i-1},\psi_{i},\Codon_{i})}{\Npsi-1},
\end{eqn}
for a given codon sequence $\Codon$ and amino acid sequence $\psi$. 
Therefore, a high CPB index indicates that the sequence uses more of the preferred codon pairs and less of the underrepresented ones.

In order to incorporate both CAI and CPB soft constraints in our method, we applied the following free energy bonus:
\begin{eqn}
    \Bias^\mathrm{CAI/CPB}_\psi(\phi) \Eq \Bias^\mathrm{CAI}_\psi(\phi) -(\Npsi-1) \log\mathrm{CPB}(\psi, \phi),
\end{eqn}
again ensuring a formulation which is extensive with respect to sequence length.

\subsubsection{Soft constraint parameters}
For CAI, the codon weights were taken from the fitness values given by \cref{eq:cai-fitness-value}, calculated using the observed codon frequencies found on \url{https://www.genscript.com/tools/codon-frequency-table}. 
For CPB, the codon pair weights were computed using the argument of the logarithm in \cref{eq:cps-def}, for which the observed frequencies for codons, codon pairs, amino acids, and amino acid pairs were required. 
We used the same observed frequencies for individual codons as used in the weights for CAI, and obtained the observed codon pair frequencies from CoCoPUTs \cite{Athey2017-new,Alexaki2019-codon} (\url{https://dnahive.fda.gov/dna.cgi?cmd=codon_usage&id=537&mode=cocoputs}). 
Each amino acid frequency was then computed by adding the frequencies of all codons encoding it, and, similarly, each amino acid pair frequency was calculated by adding the frequencies of all codon pairs encoding the amino acid pair. 
The frequencies for E. coli were used for Lipomycin polyketide synthase (LipPKS), while the frequencies for homo sapiens were used for all the other example proteins in this work.

\subsection{Tensor train operator encoding \label{si:operators}}

\newcommand{\OpList}{\vec{\tau}}
In our conception, we consider it simplest to construct the RNA sequence tensor train in the following format.
We represent each RNA sequence index $i$ by a ``operator list'' $\OpList_i$ of tuples, each of which is of the form $(a, b, c, x)$, where 
\begin{enumerate}
	\item $a \in \UpTo{\DimTT_i}$ is the left (5$'$) transfer index
	\item $b \in \mathbb{N}$ is the RNA base index (which we can encode, e.g., as one of $\lrb{0, 1, 2, 3}$)
	\item $c \in \UpTo{\DimTT_{i+1}}$ is the right (3$'$) transfer index
	\item $x \in \mathbb{R}^+$ is the scale factor, reflecting either a bonus or else 1, if there is no bonus to be applied.
\end{enumerate}
Implicitly, this provides a sparse representation of the tensor $T_i$ in which:
\begin{eqn}
	T^{a,b,c}_i = \sum_{(a',b',c',x) \in \OpList_i} \delta_{a,a'} \delta_{c,c'} \delta_{b,b'} x,
\end{eqn}
and we avoid computing the dense form of $T$, as $T$ is often fairly sparse in practice.
For minimum free energy computations, we replace the entirety of each $T$ tensor via the entrywise transformation $x' \gets -\log x$.

\subsubsection{Implementation of codon adaptation index (CAI) soft constraints}

We describe below our single codon formulation, designed to implement either CAI soft constraints or no soft constraints at all.

\begin{enumerate}
    \item For a sequence position $i$ corresponding to the 5$'$ base in a codon, $\OpList_i$ comprises $(0, b, b, 1)$ for the index $b$ of each possible base at this position, given the desired amino acid.
    \item For a sequence position $i$ corresponding to the middle base in a codon, $\OpList_i$ comprises $(a, b, c, x)$ with CAI bonus $x$ (or 1, if CAI soft constraints are not applied), index of possible 5$'$ base $a$, index of possible 3$'$ base $c$, and $b$ is the middle base for that codon, for each codon that can code for the desired amino acid.
    \item For a sequence position $i$ corresponding to the 3$'$ base in a codon, $\OpList_i$ comprises $(b, b, 0, 1)$ for the index $b$ of each possible base at this position, given the desired amino acid.
\end{enumerate}

\subsubsection{Implementation of codon pair bias (CPB) soft constraints}

We describe below our codon pair formulation, designed to implement either CPB soft constraints or both CAI and CPB soft constraints.
To combine CAI and CPB soft constraints, we simply use the CPB scheme modified such that the middle base in a codon is bonused by the proper CAI value.
\begin{enumerate}
    \item For a sequence position $i$ corresponding to the 5$'$ base in a codon, $\OpList_i$ comprises $(a, b, c, 1)$ for each possible pair of codons at positions $i$ and $i-1$, where $c$ is the index of the codon at position $i$, $a$ is the index of the codon at position $i-1$, $b$ is the base implied by the given codon at sequence position $i$, and $x$ is the CPB bonus for that codon pair.
    For the edge case of the first amino acid in the sequence, we omit the CPB bonus and insert a fictitious prior amino acid which is coded for by only one codon of index 0. 
    \item For a sequence position $i$ corresponding to the middle base in a codon, $\OpList_i$ comprises $(c, b, c, x)$ for each possible codon at that position, where $c$ is the index of that codon, $b$ is the base implied by that codon at sequence position $i$, and $x$ is either the CAI bonus for that codon, or else 1 if CAI bonuses are not being applied.
    \item For a sequence position $i$ corresponding to the 3$'$ base in a codon, $\OpList_i$ comprises $(c, b, c, 1)$ for each possible codon at that position, where $c$ is the index of that codon, and $b$ is the base implied by that codon at sequence position $i$.
\end{enumerate}

\subsubsection{Handling start and stop codons \label{si:start-stop}}
In our approach, we design only the interior codons in a specified coding sequence.
Specifically, we remove the start and stop codons ahead of time from the sequence and consider the (hypothetical) RNA strand coding only for the input amino acid sequence.
While the start and stop codons can be trivially included into our methodology, existing literature \cite{Zhang2023-algorithm} indicates that the structural stability objective should omit the start and stop codons from consideration.

\subsubsection{Handling noncoding regions \label{si:noncoding}}
For simplicity, we do not consider any noncoding regions in the present work (subsequences of an mRNA that do not directly code for an amino acid).
However, we mention here the possibility of including noncoding bases into our sequence design.
To include such bases, one needs only to use the operator list comprising $(0, b, 0, x)$ for each possible base $b$ at the given position and bonus $x$ (with $x=1$ if there is no relevant bonus to apply).

\subsection{Computational cost of codon constraints \label{si:codon-cost}}

Recall that $\DimTT_i$ is the dimension of the RNA sequence tensor train between tensors $i-1$ and $i$ (i.e., the number of values that $\sigma_i$ can take).
By inspection, our free energy algorithm runs in time complexity $O(\TotalDim^3)$ and space complexity $O(\TotalDim^2)$, where $\TotalDim = \sum_{i \in \UpTo{n}} \DimTT_i$.
For most real-world coding problems, $\TotalDim$ is lower for the single codon tensor train, making it more efficient.
However, the codon pair tensor train formulation allows for the incorporation of codon pair biases, making it moderately more flexible.
In general, the dimensionality (and cost) of the optimal tensor train increases with (1) the amount of sequence information needed to evaluate the desired set of soft constraints and (2) the locality of that information with respect to sequence position.
Common soft constraints (e.g., CAI and CPB) are, nicely, naturally localized and relatively minimal in this respect.
(We discussed a general way of formulating tensor trains for a specified set of sequence constraints in \cref{si:operators}.)

\section{Pseudocode and algorithmic details \label{si:pseudocode}}

\subsection{Partition function computation \label{si:pf}}


First, in \cref{alg:complex-pf}, we give our main method for the computation of the free energy $\qpsi$.
This algorithm is nearly the same as its corresponding algorithm in \cite{Fornace2025-new}, except that it replaces the folding function $\Fold$ by the folding function $\DesignFold$, which incorporates the sequence variability yielded by an input amino acid sequence $\psi$.
Note that our presentation relies on the consistent flattening of indices, i.e. treating each $Q_{i,j}$ tensor as a matrix (grouping its indices like $Q_{i,j}^{\sigma_i, \zeta_i, \sigma_j, \zeta_j} \Rightarrow Q_{i,j}^{(\sigma_i, \zeta_i), (\sigma_j, \zeta_j)}$)
This accomplishes not just a summation over secondary structures but also a summation of all feasible RNA sequences.

\begin{algorithm}[H]
    \caption{$O(n^3)$ algorithm to calculate $\qpsi$}
    \label{alg:complex-pf}
    \begin{algorithmic}
        \Function{AbstractFreeEnergy}{$n, \DesignFold$}
            \For{$j \in \UpTo{\Nphi}$}  
                \ForAll{$i \in \UpTo{j+1}$} 
                    \State $Q_{j, i} \gets \DesignFold_{i,j}(Q_{i+1,j-1})$
                \EndFor
                \ForAll{$i \in \UpTo{j}$} 
                    \State $Q_{i,j} \gets \sum_{k \in \Range{i}{j}} Q_{i, k} Q_{j, k}^\t$ 
                \EndFor
            \EndFor
            \State \Return $(Q, \; \ologq(n, \DesignFold, Q))$
        \EndFunction
        \\
        \Function{FreeEnergy}{$\psi$}
            \State $\DesignFold \gets$ folding function derived from $\psi$ and secondary structure free energy model
            \State $(Q, \qpsi) \gets$ \Call{AbstractFreeEnergy}{$3 \cdot \Abs{\psi}, \DesignFold$}
            \State \Return $\qpsi$
        \EndFunction
    \end{algorithmic}
\end{algorithm}

For reference, we additionally give a more efficient algorithm for free energy computation \cref{alg:fast-pf} using matrix primitives, which may replace \textsc{AbstractFreeEnergy} above.
This algorithm is essentially the same as the corresponding one from \cite{Fornace2025-new}, replacing $\Fold$ with $\DesignFold$.

\begin{algorithm} [H]
    \caption{$O(n^3)$ algorithm to calculate $\qpsi$ using matrix primitives }
    \label{alg:fast-pf}
    \begin{algorithmic}
        \Function{EfficientFreeEnergy}{$n,\DesignFold$}
            \For{$j \in \UpTo{\Nphi}$}
                \State [If using GPUs and $j>1$, get $Q_{\Range{1}{j}, j-1}$ from GPUs] \Comment{$O(n \Dim^2)$ data transfer}
                \ForAll{$i \in \UpTo{j+1}$} 
                    \State $Q_{j, i} \gets \DesignFold_{i,j}(Q_{i+1,j-1})$ \Comment{Folding}
                \EndFor
                \State $\Call{Renormalize}{Q, x, j}$ \Comment{Renormalization}
                \State [If using GPUs, send $Q_{j, \Inclusive{0}{j}}$ to GPUs] \Comment{$O(n \Dim^2)$ data transfer}
                \State $Q_{\UpTo{j}, j} \gets \Call{Trmm}{Q_{\UpTo{j}, \UpTo{j}}, Q_{j, \UpTo{j}}}$ \Comment{Propagation, optionally on GPUs}
            \EndFor
            \State \Return $(Q, \ologq(n, \DesignFold, Q))$
        \EndFunction
    \end{algorithmic}
\end{algorithm}
Here, \textsc{Trmm} is the triangular matrix multiplication:
\begin{eqn}
    (\textsc{Trmm}(A, B))_{i,j} = \sum_{k=i}^n A_{i,k} B_{j, k}
\end{eqn}
for a square matrix $A$ of size $n,n$ and rectangular matrix $B$.
(In this routine, $Q$ is viewed in its flattened matrix form.)
Meanwhile, \textsc{Renormalize} is a routine renormalizing each row $Q_{j,\UpTo{j}}$ to avoid floating point overflow (cf., \cite{Fornace2025-new}).
In this work, in particular, we sought a normalization for each row such that the maximum spectral radius of each block was $\approx$1 (padding each rectangular matrix with zeros beforehand as needed).
From this intuition, we approximated this factor in each iteration as:
    $f = \max_i (A x)_i / x_i$,
with $A = Q_{j,\UpTo{j}}^\t$ and $x$ a single random vector with entries drawn i.i.d. and uniformly from $[0,1]$.
This heuristic proved effective in all of our investigations at preventing any partition function overflow for large sequences.

For this work, we implement multi-GPU acceleration for the first time.
For this approach, we take advantage of the more detailed block definition of $Q$ defined in \cite{Fornace2025-new}.
In essence, each $Q_{i,j}$ has block structure:
\begin{eqn}
    Q_{i,j} = \begin{bmatrix}
        Q^{(0,0)}_{i,j} & { } & { } & { } & { } \\
        Q^{(1,0)}_{i,j} & Q^{(1,1)}_{i,j} & { } & { } & { } \\
        { } & { } & Q^{(2,2)}_{i,j} & { } & { } \\
        { } & { } & { } & \ddots & { } \\
        { } & { } & { } & { } & Q^{(\blocks,\blocks)}_{i,j}
    \end{bmatrix}
\end{eqn}
Therefore, we distribute the blocks of $Q$ across available GPUs, with the constraint that blocks $Q^{(0,0)}$, $Q^{(1,0)}$, and $Q^{(1,1)}$ are on the same GPU.
This avoids any need for any direct synchronization between different GPUs within the propagation (\textsc{Trmm}) operation.


\subsection{Reverse interval computation \label{si:reverse}}

We next give our algorithm for the computation of complementary ``reverse'' free energies, needed to compute base pair probabilities via \cref{eq:pair-formula} and marginal sequence probabilities via \cref{eq:marginal-base,eq:marginal-codon}.
In essence, while the forward dynamic programs of \cref{si:pf} compute the partition functions of all subsequences $\Range{i}{j}$ for $0 \leq i \leq j < n$, \cref{alg:pairs} computes the partition functions of all subsequences $\Range{j}{i+n}$ for $0 \leq i < j < n$, where the range now cycles back to the beginning (i.e., for example, $i=3$, $j=2$, and $n=5$ would refer to the sequence $\phi_{j,i+n} = (\phi_{3},\phi_4,\phi_0, \phi_1)$). 
This is essentially the same as that from \cite{Fornace2025-new}, replacing $\Fold$ with $\DesignFold$.

\input{pseudocode/pairs.tex}

\Cref{alg:pairs} may also be parallelized and GPU-accelerated in an approach completely analogous to \cref{alg:complex-pf}.

\subsection{Simultaneous sequence sampling \label{si:sample-alg}}

For fixed RNA sequences, direct Boltzmann sampling of secondary structures has been a standard algorithm since \cite{Ding2003-statistical}.
In \cite{Fornace2020-unified}, we extended this approach to multistranded complexes and stacking recursions, while we also proposed a \emph{simultaneous sampling} algorithm, which achieves speedups for the common scenario in which a user requests multiple i.i.d. sampled structures.
In \cite{Fornace2025-new}, we extended this approach even further, improving our algorithm and generalizing it to tensor-based free energy models, which yielded large simplifications.
In particular, the respective algorithm in \cite{Fornace2025-new} exploits and improves the acceleration of simultaneous structure sampling \cite{Fornace2020-unified}.
In that prior work, we noted that in most natural and engineered systems, independent structure sampling would concentrate strongly to yield around, if not just a single structure, at least a combination of a small number of recurrent substructures.
Algorithmically, this leads to the common revisiting of recursion elements over and over if deriving samples in a sequential manner.
In fact we can use the entirety of the approach developed there.


In this section, we give a full algorithm for simultaneously sampling $J$ i.i.d. combinations of RNA sequences and secondary structures.
This algorithm builds upon that of \cite{Fornace2025-new}, such that each yielded state contains both (1) secondary structure (base-pairing) information and (2) a list of RNA sequence tensor train operator indices.
From the latter outputs, it is simple to reconstruct the sampled RNA sequence (as well as the specific codon indices and incurred soft constraint bonuses).
See \cref{si:operators} for details of how we construct the RNA sequence tensor train in terms of individual operators.

In \cref{alg:sample}, for simplicity, the core sampling routine $\textsc{SampleAlternatives}$ is abstracted, taking a list of alternative states, each of which contains (1) a set of recursion elements, (2) a set of operator indices, and (3) an associated partition function.
In practice, the list of alternative states need not be constructed eagerly.
In addition, a few quantities involving $B$ terms may be precontracted ahead of time, for some mild cost savings.

\input{pseudocode/sample.tex}

In \cref{alg:sample}, $\textsc{Operators}$ is a function which, supplied an RNA sequence index, returns an operator list (\cref{si:operators}) of tuples for the respective tensor at that position in the RNA sequence tensor train.
Each is of the form $(z,a,b,c,x)$, with $z$ the index of the tuple and $a,b,c,x$ reflecting the $5'$ index, RNA base, $3'$ index, and multiplicative bonus (respectively) as laid out in \cref{si:operators}.

In more detail, $\Queue$ is a priority queue built upon a partial ordering of recursion elements.
This ordering is constrained such that $E_1 < E_2$ for any recursion element $E_1$ depending on another recursion element $E_2$ (indirectly or directly) via the recursion equation \cref{eq:dp-short}. 
(See \cite{Fornace2025-new} for details of this ordering, which operates identically in this work.)
The alternative state set $\vec{v}$ is pre-sorted so that the sampled structures can be distributed efficiently using a binary search routine \textsc{UpperBound}.

\FloatBarrier
\subsection{Suboptimal sequence generation \label{si:subopt}}

In this section, we give a full algorithm for reconstructing suboptimal combinations of RNA sequences and secondary structures.
This algorithm builds upon that of \cite{Fornace2025-new}, such that each yielded state contains both (1) secondary structure (base-pairing) information and (2) a list of RNA sequence tensor train operator indices.
From the latter outputs, it is simple to reconstruct the optimized RNA sequences (as well as the specific codon indices and incurred soft constraint bonuses).
See \cref{si:operators} for details of how we construct the RNA sequence tensor train in terms of individual operators.


\input{pseudocode/subopt.tex}

The function $\textsc{Operators}$ operates as in \cref{si:sample-alg}, except that the soft constraint bonus is returned as a free energy rather than a multiplicative bonus (by taking the negative logarithm of the latter).
The ordered map $\Queue$ uses the same key ordering as in \cref{si:sample-alg} as well.
To compute all sequences solving the minimum problem \cref{eq:min-problem}, it suffices to invoke \cref{alg:subopt} with $\gap = 0$.
To compute just one sequence solving that problem, it suffices to invoke \cref{alg:subopt} with $\gap < 0$.
In the main iteration \cref{alg:subopt}, the state indices $\Ix$ are sorted so that the suboptimal states can be determined efficiently using a binary search routine \textsc{LowerBound}.

\FloatBarrier
\subsection{A simple iterative approach to frequency matching \label{si:iterative-reweighting}}

In this section, we introduce a simple approach to frequency matching in our methodology.
To motivate this algorithm, recall that the central purpose of CAI and CPB soft constraints is to yield designed RNA sequences which match the codon or codon pair frequencies of the target host.
Implementing CAI soft constraints as we have laid out, therefore, provides a simple and direct sampling algorithm which approximately matches those frequencies, at least if the unbonused sampling distribution $\Dist(\psi)$ is roughly uniform with respect to sequence $\phi$.
However, this heuristic is only approximate and only qualitatively approximates these desired frequencies in practice (i.e., undesirably underweighting or overweighting different codons).

Rather than necessitating arbitrary user-chosen scalings of sequence-based soft constraints, we show here that matching desired sampling frequencies may be accomplished by means of a short iterative procedure.
Namely, the \emph{naive} version of our procedure assumes the following iteration:
\begin{enumerate}
    \item Given the current soft constraint bonus values, exactly calculate the output codon or codon pair frequencies of our sampling distributions by summing the exact marginal probabilities over all sequence indices
    \item Multiply the current soft constraint bonus values by the ratio of desired frequencies divided by the achieved frequencies
    \item Repeat until convergence
\end{enumerate}
This fixed-point iteration can be formalized in the following algorithm:
\begin{algorithm}[H]
    \caption{Naive iterative algorithm for frequency matching} 
    \label{alg:iterative}
    \begin{algorithmic}
        \Function{MatchFrequencies}{$\psi, f_*, r_{\mathrm{tol}}$} \Comment{Match desired frequencies $f_*$ up to $\ell_\infty$ tolerance $r_\mathrm{tol}$}
            \State $b_* \gets$ computed frequencies from $\mathcal{U}(\psi)$ with applied multiplicative bonuses $f_*$
            \State $f \gets f_*$
            \While{true}
                \State Compute $Q$ and free energy $\qpsi$ using bonuses $f$
                \State Compute marginal probabilities from $Q$ and $\qpsi$ (e.g., using \cref{eq:codon-marginal} for single codon marginals)
                \State $b \gets $ accumulated codon or codon pair frequencies from these marginal probabilities
                \If{$\Norm{\log(b / b_*)}_\infty < r_\mathrm{tol}$} 
                    \textbf{break}
                \EndIf \Comment{entrywise division and logarithm}
                \State $f \gets f \cdot b / b_*$ \Comment{entrywise multiplication and division}
            \EndWhile
            \State \Return $f$
        \EndFunction
    \end{algorithmic}
\end{algorithm}
\noindent In the above, we omit considerations of any requested frequencies $b_*$ equal to zero.

In practice, we observe that much faster convergence can often be assured by applying a standard implementation of Anderson acceleration to \cref{alg:iterative} \cite{Walker2011-anderson}. 
In \cref{fig:frequency-matching}, we show that good convergence can be obtained in 8-15 iterations across each of our example systems, regardless of system size.
After the procedure is finished, sequence sampling may be performed with the output bonuses, matching the desired frequencies nearly exactly in expectation (though each individual sampled sequence will likely only approximately match the desired frequencies).

\input{./floats/frequency-matching.tex}

\subsection{Computational details \label{si:computational-details}}

\label{si:parallelization}

All computational benchmarks were run on the NERSC Perlmutter cluster (CPU: AMD EPYC 7763, GPU: Nvidia A100) and used single precision floating point operations.
For \cref{fig:performance}, the ``Parallel CPU / 2$\times$GPU'' calculation used up to 64 CPU cores and 2 GPUs.
To parallelize across GPUs, we exploited the block formulation of matrices $V$ and $B$ from \cite{Fornace2025-new}.
In more detail, since each of those matrices is block diagonal (or nearly block diagonal), the matrix multiplications in \cref{alg:complex-pf} can be split by block into different GPUs.
In the utilized $(\blocks=2,\blocksize=3,\Np=1)$ model, there are 4 blocks (2+2, with the additional two serving to prevent disconnected structures from consideration).
The first two of these must be kept in the same device's memory, owing again to the way in which disconnected structures are prevented.
Therefore, in the 2$\times$GPU calculation, the first two blocks of $Q$ were calculated on GPU 1 and the second two blocks on GPU 2 (with $\DesignFold$ operations and synchronizations still handled on CPU, in parallel).

For the example studies (\cref{fig:studies}, \cref{tab:runtime-table}), each calculation of $\qpsi$ and reverse matrix $R$ was parallelized across 2 GPUs, using up to 64 CPU cores for the $\DesignFold$ folding operations.
The output matrices $Q$ and $R$ were then copied to CPU memory.
(This copy operation took a maximum of 6\% of the time as the cost of calculating $Q$ and $R$.)
Boltzmann sampling, marginal codon probabilities, and equilibrium base pair probabilities ($P^\psi$) were then calculated in serial, all incurring a much lower cost than the prior calculation of $Q$ and $R$.
After sampling 2500 RNA sequences $\phi$, complex free energies $\qpsi(\phi)$ and equilibrium base pair probabilities $(P^\phi)$ were then computed in parallel using the same CPU/GPU setup.
Given any substantial number of sampled sequences, evaluating $\qpsi(\phi)$ to construct histograms like those in \cref{fig:studies} is more expensive than running than computing $\qpsi$ and the sampled sequences themselves.
On the other hand, these computations are embarrassingly parallelizable, and we plan in the future to optimize them further via caching of shared subsequences.

Minimum free energy (MFE) calculations were performed using up to 64 CPU cores without GPU acceleration.
In the future, we plan to implement CUDA kernels for the respective matrix operations, mirroring those in CuBLAS which enable partition function calculations.
This would likely speed up our implementation of MFE calculations significantly.
However, it is worth pointing out that partition function calculations (and therefore the sampling approach introduced in this paper) are altogether better targets for high performance computing, as significantly more investment has been made to optimize matrix multiplication operations compared to minimum summation operations.
For this reason, we expect that the cost of computing $\qpsi$ will always be lower than that of computing the minimum free energy.

\label{si:benchmark}
For the benchmark of \cref{fig:performance}, we generated amino acid sequences by randomly choosing each amino acid with weight proportional to its frequency in System~3 (SPIKE\_SARS2).
Corresponding RNA sequences (for \cref{fig:performance}b) were generated by randomly selecting a codon for each amino acid with equal probabilities for each codon. 
Timings represent single replicates, as timing differences between randomly sampled sequences were observed to be qualitatively negligible.

\section{Example studies \label{si:studies}}

\subsection{Sequence and parameter details \label{si:study-details}}

We use the $(\blocks=2,\blocksize=3,\Np=1)$ free energy model regressed in \cite{Fornace2025-new}, itself derived via Kullback-Leibler divergence minimization with respect to the \texttt{rna06} nearest-neighbor model.
Alternative models were derived in that work to fit experimental melt data directly.
While those models might be more accurate, we chose the KL-derived model to be more conservative, and better validated given the long history of nearest-neighbor models.
The chosen model replicates the structure ensemble of the \texttt{rna06} nearest-neighbor model to approximately 10\%; see \cite{Fornace2025-new} for more details.

For our example studies, we evaluated the wild-type RNA sequences from the following sources:
\begin{itemize}\setlength\itemsep{0em}
    \item System 1 used the homo sapiens wild-type sequence from \url{https://www.ncbi.nlm.nih.gov/nuccore/U87459}.
    \item System 2 used the homo sapiens wild-type sequence from \url{https://www.ncbi.nlm.nih.gov/nuccore/NM_007265.3}.
    \item System 3 used the Wuhan-Hi-1 SARS-CoV-2 isolate from \url{https://www.ncbi.nlm.nih.gov/nuccore/NC_045512.2}. 
    \item System 4 used the E. coli wild-type sequence from \url{https://public-registry.jbei.org/entry/21267}
\end{itemize}
As mentioned in \cref{si:start-stop,si:noncoding}, we omit start codons, stop codons, and noncoding regions from consideration in our optimization.


\subsection{Example results including CAI and CPB bonuses}

In this section, we display sampling histograms for each of our studies, including variants (1) with sequence-based bonuses from CAI and (2) with sequence-based bonuses from CAI and CPB.

As in the main text, we include the histograms resulting from performing sampling according to a simplified $(\blocks=1,\blocksize=1,\Np=1)$ tensor model, which is essentially identical to a model which gives each type of base pair (A$\cdot$U, C$\cdot$G, G$\cdot$U) a flat free energy bonus.
This simplistic pairing model is built using only one interior block of size (1, 1) and a pairing rank of 1, the minimal case to correctly handle strand breaks in our tensor methodology.
This is to be compared to our proposed $(\blocks=2,\blocksize=3,\Np=1)$ model, which has two interior blocks of size $(3, 3)$ with a pairing rank of 1.
We evaluate all output RNA sequences using the more advanced $(\blocks=2,\blocksize=3,\Np=1)$ model.
In this section, we refer to the sampling distribution of the simplistic pairing model as $\mathcal{D}^0(\psi)$ and the MFE sequence of that model as $\phi^0_{\mathrm{min}}$.

\begin{figure}[H]
    \centering
    \includegraphics[width=0.96\textwidth]{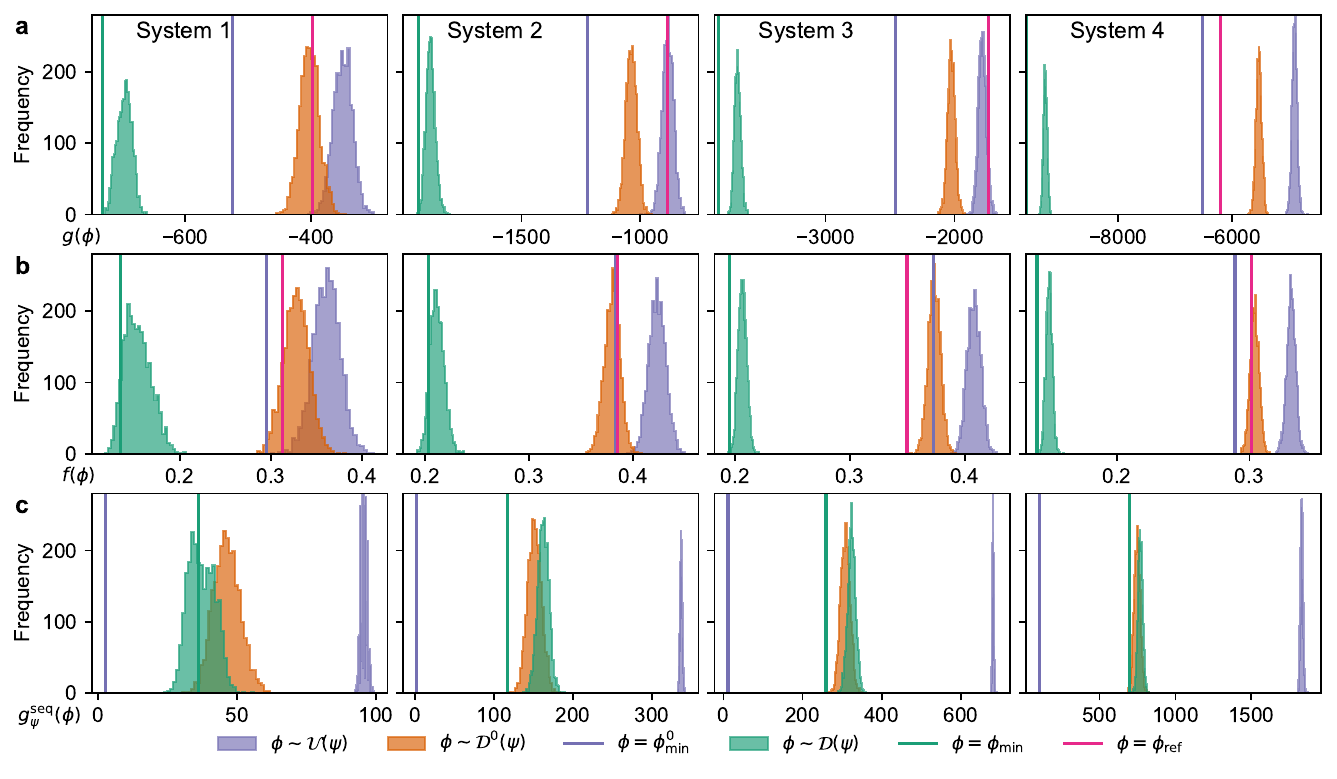}
    \vspace{-0.15in}
    \caption{\label{fig:studies-cai} Results for example systems listed in \cref{s:studies} (1-4) with CAI soft constraints applied.
    \textbf{a}: physical complex free energy $\logq(\phi)$ (unitless; multiply by 0.62 kcal/mol for results at 37\textdegree C).
    \textbf{b}: equilibrium fraction of unpaired bases $f(\phi)$.
    \textbf{c}: sequence-based free energy (from CAI soft constraints) $\qseq_\psi(\phi)$.
    Each plot depicts the different probability density distributions which can be sampled from using our provided algorithms.
    $\phi \sim \mathcal{U}(\psi)$ was generated by sampling 2500 codon sequences independently (with no consideration of secondary structure).
    $\phi \sim \mathcal{D}(\psi)$ and $\phi \sim \mathcal{D}^0(\psi)$ were generated by sampling 2500 codon sequences, respectively.
    $\phi_{\mathrm{min}}$ and $\phi^0_{\mathrm{min}}$ are the optima yielded by solving \cref{eq:min-problem} using the proposed and simplistic models, respectively, as in \cref{fig:studies}.
    }
\end{figure}

We first show results for sampling with CAI bonuses in \cref{fig:studies-cai}.
We include, as a third observable, the sequence-based free energy of the sampled sequences, which includes just the sequence-based bonuses.
Again, our proposed sampling distribution achieves a satisfactory optimization of both sequence and structural objectives simultaneously.
The simplified sampling methodology performs worse in the structural objective.

\begin{figure}[H]
    \centering
    \includegraphics[width=0.74\textwidth]{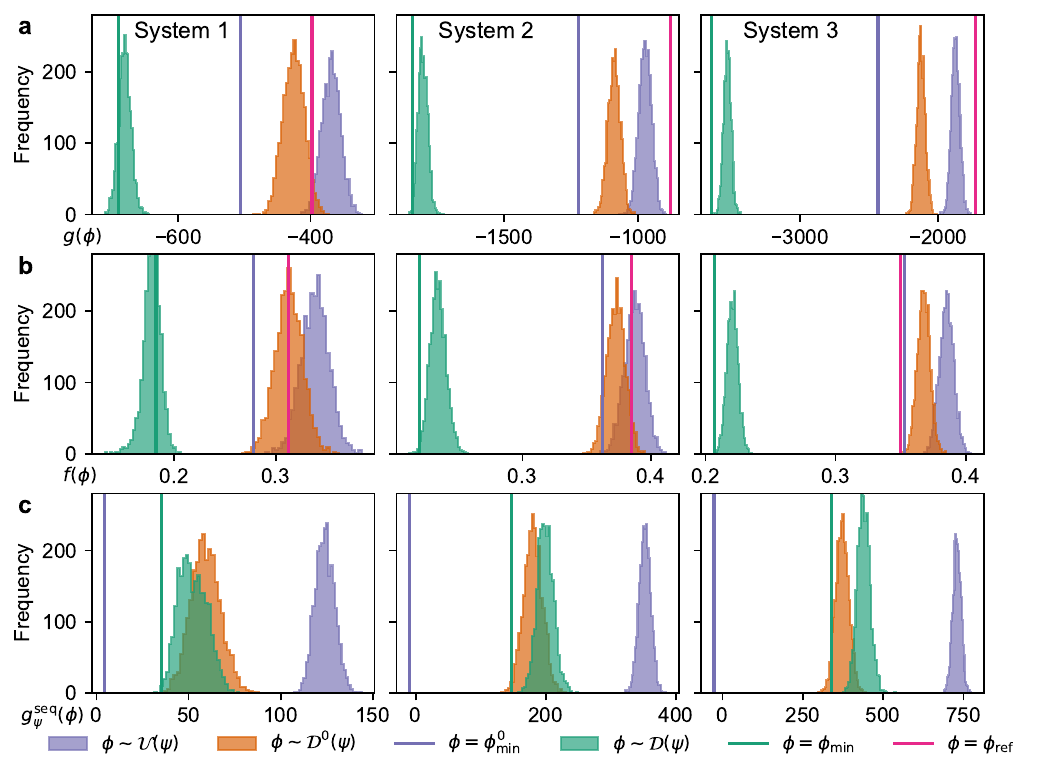}
    \vspace{-0.15in}
    \caption{\label{fig:studies-cpb} Results for example systems listed in \cref{s:studies} (1-3) with CAI and CPB soft constraints applied.
    \textbf{a}: physical complex free energy $\logq(\phi)$ (unitless; multiply by 0.62 kcal/mol for results at 37\textdegree C).
    \textbf{b}: equilibrium fraction of unpaired bases $f(\phi)$.
    \textbf{c}: sequence-based free energy (from CAI and CPB soft constraints) $\qseq_\psi(\phi)$.
    Each plot depicts the different probability density distributions which can be sampled from using our provided algorithms.
    $\phi \sim \mathcal{U}(\psi)$ was generated by sampling 2500 codon sequences independently (with no consideration of secondary structure).
    $\phi \sim \mathcal{D}(\psi)$ and $\phi \sim \mathcal{D}^0(\psi)$ were generated by sampling 2500 codon sequences, respectively.
    $\phi_{\mathrm{min}}$ and $\phi^0_{\mathrm{min}}$ are the optima yielded by solving \cref{eq:min-problem} using the proposed and simplistic models, respectively, as in \cref{fig:studies}.
    }
\end{figure}

Finally, we show results for sampling with CAI and CPB soft constraints in \cref{fig:studies-cpb}.
(In this case, we omit the largest system due to our GPU memory constraints.)
We include, as a third observable, the sequence-based free energy of the sampled sequences, which includes just the sequence-based soft constraint bonuses.
Again, our proposed sampling distribution achieves a satisfactory optimization of both sequence and structural objectives simultaneously.
The simplified sampling methodology performs worse in the structural objective.

\subsection{Example results for marginal codon probabilities and base pair probabilities}
\FloatBarrier

In this section, we showcase exactly calculated statistics of the sampling distributions for each of our example studies (excluding CAI and CPB bonuses).
For each system, we include:
\begin{itemize}\setlength\itemsep{0em}
    \item the marginal codon probabilities $\pcodon_j(c)$ for each amino acid in the target protein, as a function of position $j$.
    We display these as a heatmap, in which each column $j$ holds the computed probabilities of each codon compatible with amino acid $\psi_j$ (in an arbitrary but consistent order). 
    Each column is padded below by zeros to have an equal length.
    \item the accumulated codon frequencies for each amino acid and codon combination
    \item the equilibrium unpaired probabilities $\Ppsi_{i,i}$ as a function of RNA sequence position $i$. The considered stability objective penalizes overly unpaired sequences.
\end{itemize}

These results exemplify the rich statistics that can be exactly obtained, all in $O(n^3)$ time with CPU/GPU acceleration, using our approach.

\begin{figure}[H]
    \centering
    \includegraphics[width=0.9\textwidth]{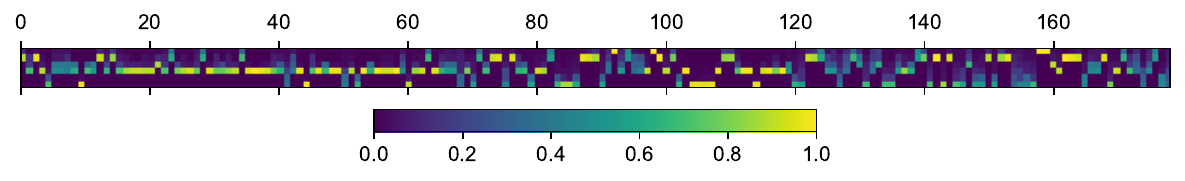}
    \caption{\label{fig:marginal-a} Marginal codon probabilities for System 1 in \cref{s:studies}.
    $\Prob^\psi_j(c)$ is depicted for each possible codon $c$ (in arbitrary vertical order) and amino acid position $j$.
    }
\end{figure}

\begin{table}[H]
    \centering
    \input{./tables/codon-frequencies-a.tex} \vspace{1em}
    \caption{\label{tab:codon-frequencies-a} \normalfont{Accumulated codon frequencies for System 1.}}
\end{table}

\begin{figure}[H]
    \centering
    \includegraphics[width=0.7\textwidth]{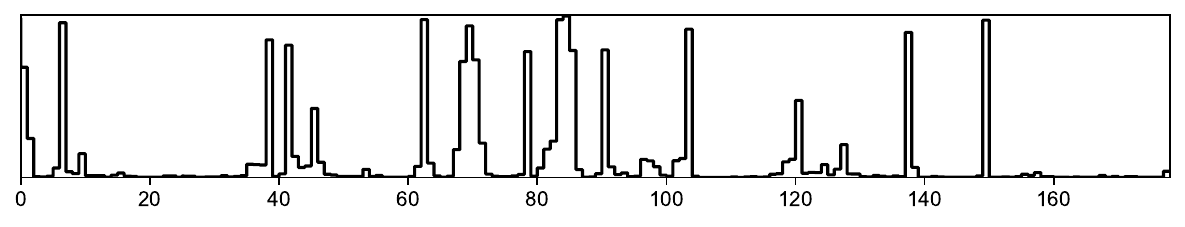}
    \caption{\label{fig:unpaired-a} Equilibrium unpaired probabilities $\Ppsi_{i,i}$ for System 1 in \cref{s:studies}.
        ($y$-axes are normalized to span [0, 1].)
    }
\end{figure}

\begin{figure}[H]
    \centering
    \includegraphics[width=0.9\textwidth]{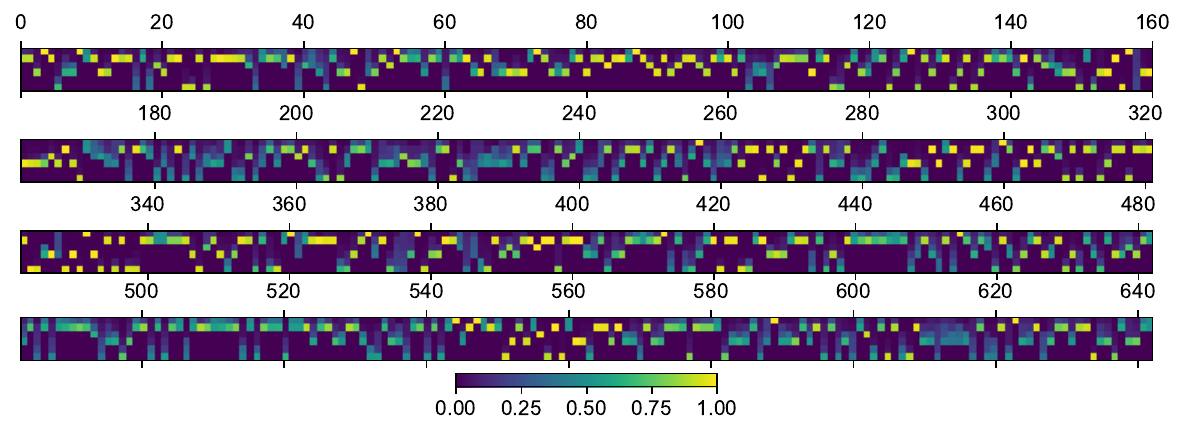}
    \caption{\label{fig:marginal-b} Marginal codon probabilities for System 2 in \cref{s:studies}.
    $\Prob^\psi_j(c)$ is depicted for each possible codon $c$ (in arbitrary vertical order) and amino acid position $j$.
    }
\end{figure}

\begin{table}[H]
    \centering
    \input{./tables/codon-frequencies-b.tex} \vspace{1em}
    \caption{\label{tab:codon-frequencies-b} \normalfont{Accumulated codon frequencies for System 2.}}
\end{table}

\begin{figure}[H]
    \centering
    \includegraphics[width=0.9\textwidth]{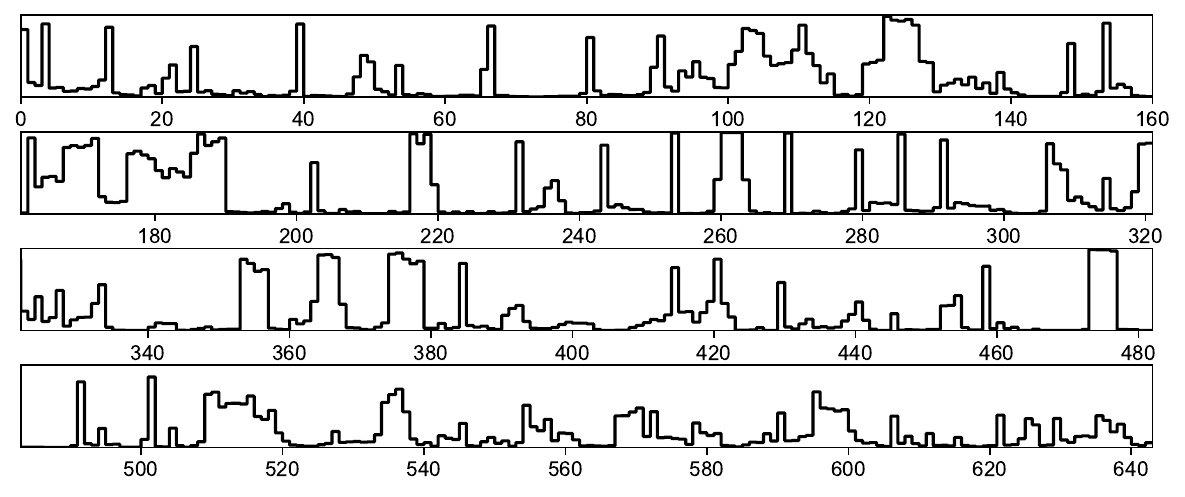}
    \caption{\label{fig:unpaired-b} Equilibrium unpaired probabilities $\Ppsi_{i,i}$ for System 2 in \cref{s:studies}.
        ($y$-axes are normalized to span [0, 1].)
    }
\end{figure}

\begin{figure}[H]
    \centering
    \includegraphics[width=\textwidth]{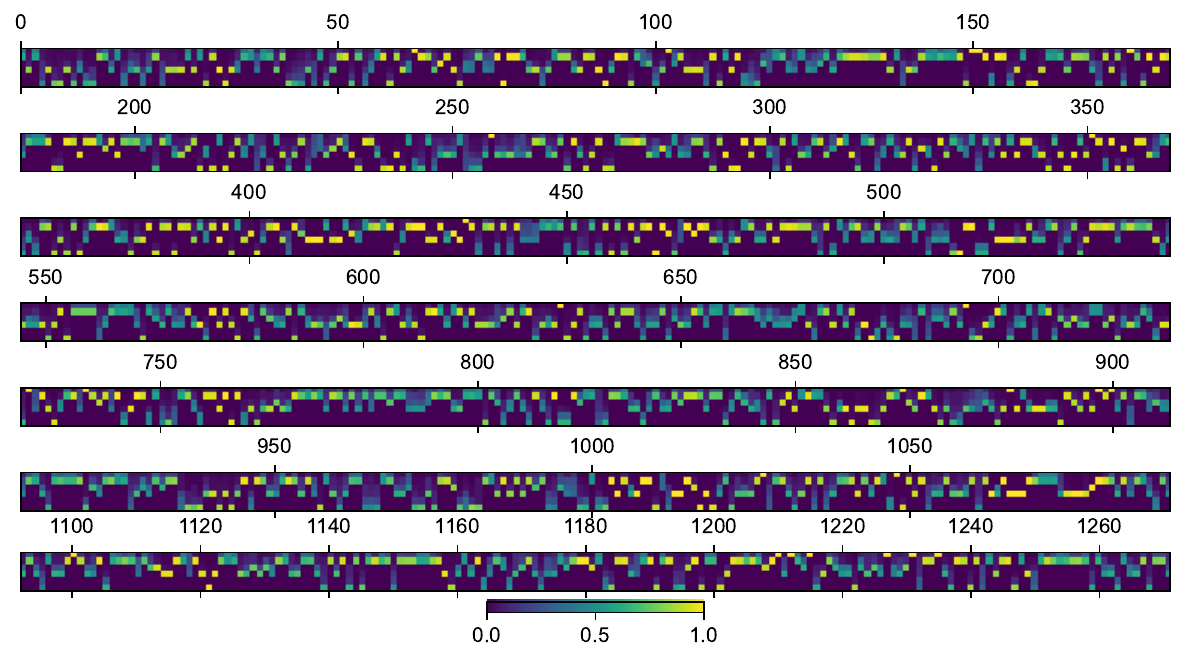}
    \caption{\label{fig:marginal-c} Marginal codon probabilities for System 3 in \cref{s:studies}.
    $\Prob^\psi_j(c)$ is depicted for each possible codon $c$ (in arbitrary vertical order) and amino acid position $j$.
    }
\end{figure}

\begin{table}[H]
    \centering
    \input{./tables/codon-frequencies-c.tex} \vspace{1em}
    \caption{\label{tab:codon-frequencies-c} \normalfont{Accumulated codon frequencies for System 3.}}
\end{table}

\begin{figure}[H]
    \centering
    \includegraphics[width=0.9\textwidth]{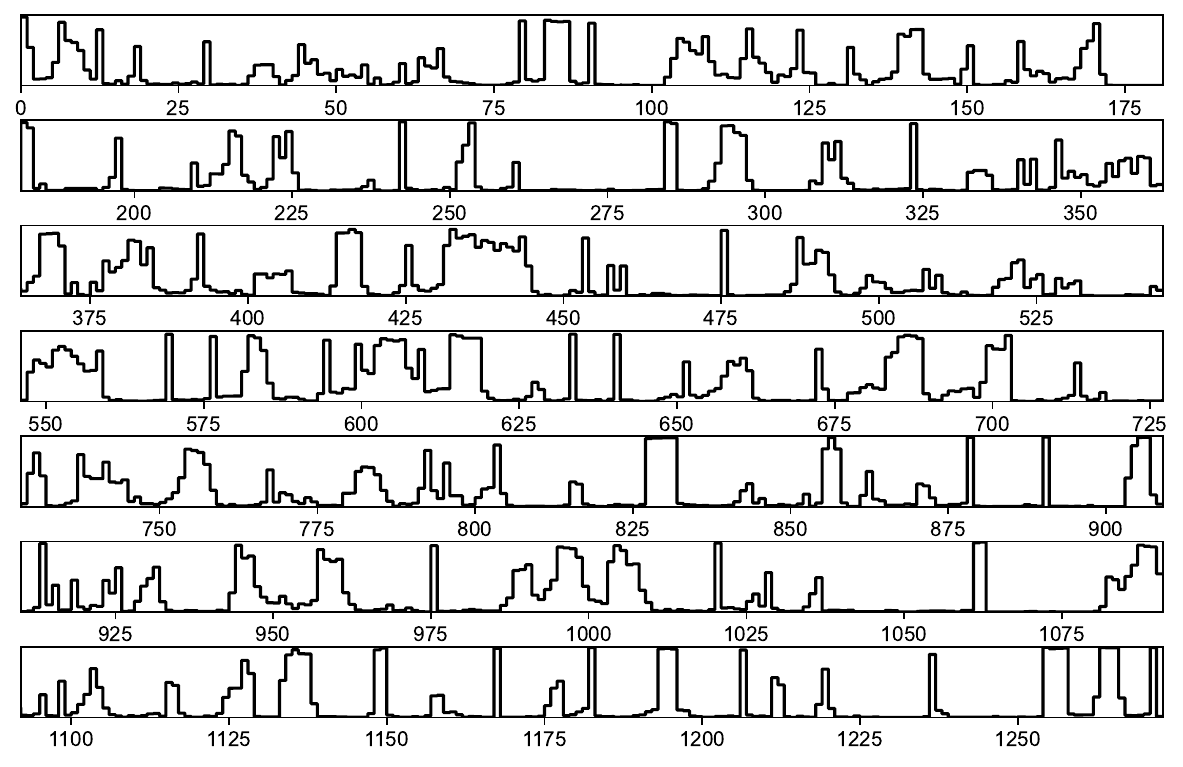}
    \caption{\label{fig:unpaired-c} Equilibrium unpaired probabilities $\Ppsi_{i,i}$ for System 3 in \cref{s:studies}.
        ($y$-axes are normalized to span [0, 1].)
    }
\end{figure}

\begin{figure}[H]
    \centering
    \includegraphics[width=\textwidth]{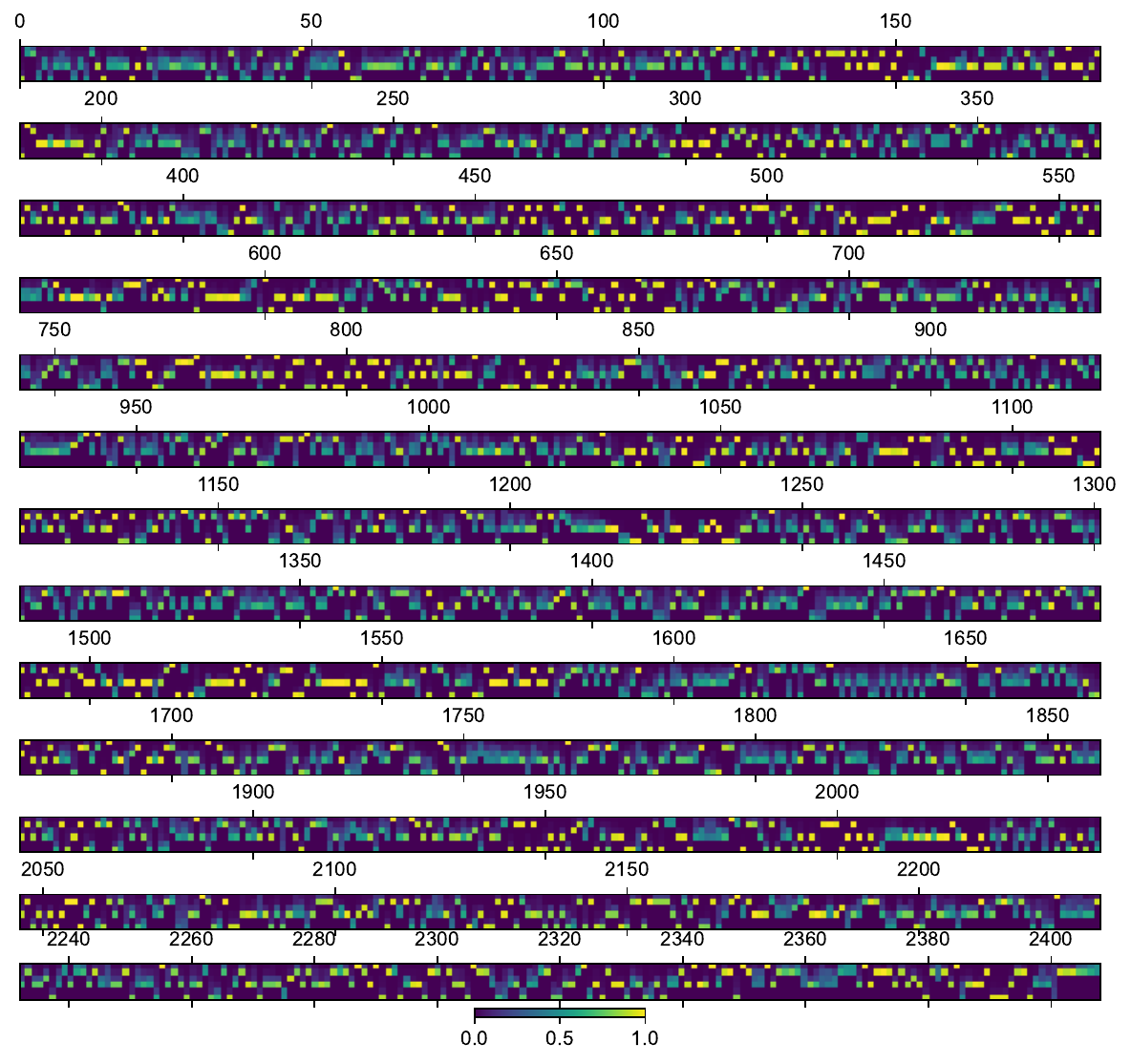}
    \caption{\label{fig:marginal-d} Marginal codon probabilities for System 4 in \cref{s:studies}.
    $\Prob^\psi_j(c)$ is depicted for each possible codon $c$ (in arbitrary vertical order) and amino acid position $j$.
    }
\end{figure}

\begin{table}[H]
    \centering
    \input{./tables/codon-frequencies-d.tex} \vspace{1em}
    \caption{\label{tab:codon-frequencies-d} \normalfont{Accumulated codon frequencies for System 4.}}
\end{table}

\begin{figure}[H]
    \centering
    \includegraphics[width=0.9\textwidth]{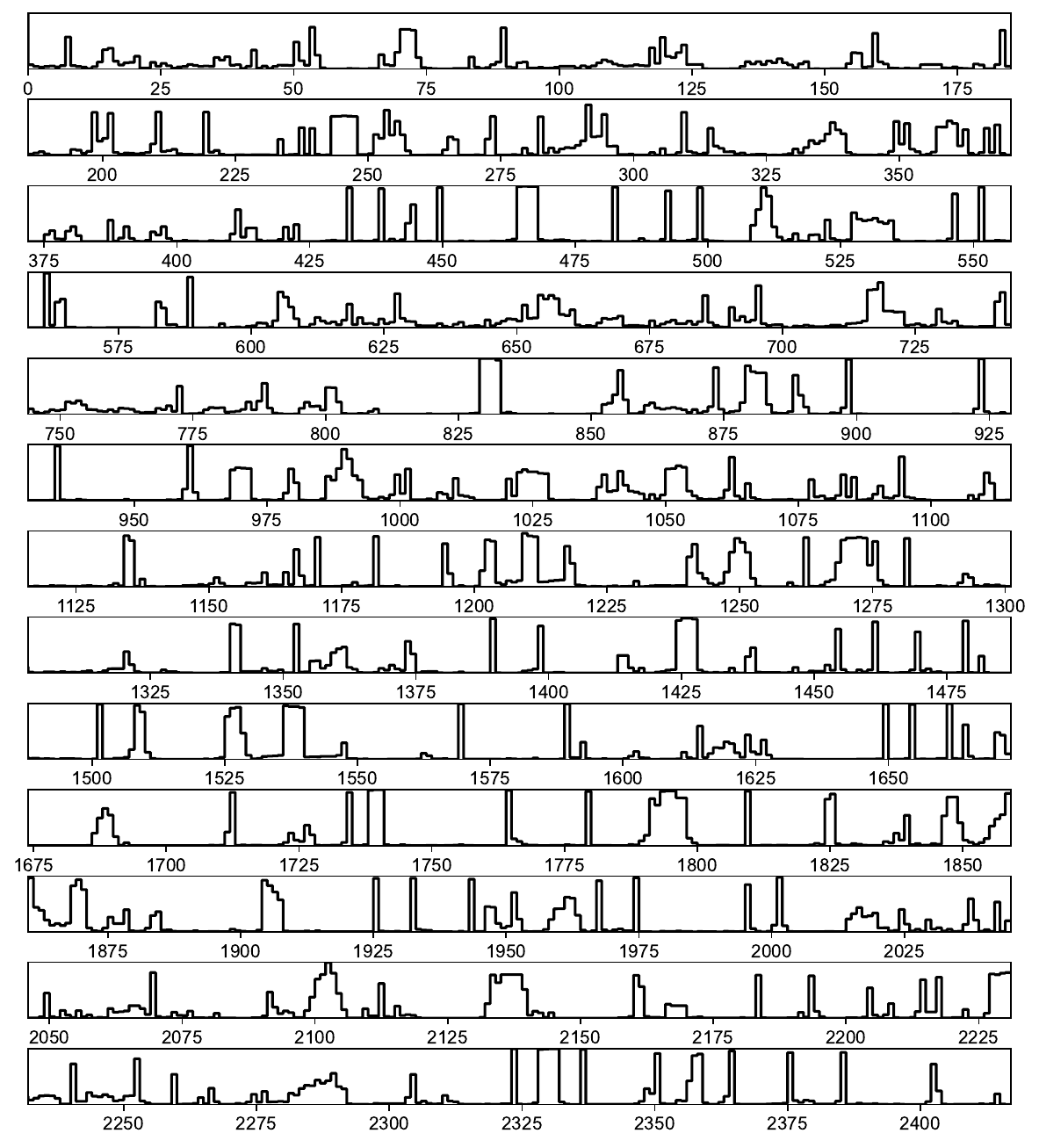}
    \caption{\label{fig:unpaired-d} Equilibrium unpaired probabilities $\Ppsi_{i,i}$ for System 4 in \cref{s:studies}.
        ($y$-axes are normalized to span [0, 1].)
    }
\end{figure}

\section{Alternative method comparisons}
\label{si:alternative-methods-comparisons}

In this section, we discuss the most closely related alternative methods in more depth, namely (1) LinearDesign\cite{Zhang2023-algorithm}, (2) EnsembleDesign\cite{Dai2025-EnsembleDesign}, and (3) CDSFold\cite{Terai2015}.
We provide performance benchmarks for each method.
As noted before, these methods vary substantially from our approach, including different optimization objectives, different algorithmic approaches, and different underlying free energy models.
Therefore these benchmarks only yield a rough guide to understand computational costs and tradeoffs.

More specifically, in contrast to CDSfold and LinearDesign, which optimize a single minimum-free-energy structure, and EnsembleDesign, which represents discrete codon choices as probabilities and applies gradient optimization with beam-pruned probabilistic lattice parsing to approximate an ensemble-based objective, our method directly defines and samples from a Boltzmann distribution over all synonymous coding sequences and their compatible secondary structures. Our tensor-based dynamic-programming formulation enables exact sampling from the specified Boltzmann distribution and computation of ensemble statistics. 
Finally, our approach can also incorporate codon-based objectives and supports parallel computation on CPUs and GPUs.
(The publicly available codes for the alternative methods are all CPU-based, without GPU implementations.)

\subsection*{LinearDesign} First, LinearDesign \cite{Zhang2023-algorithm} selects an mRNA sequence encoding a prescribed protein by jointly optimizing minimum folding free energy (MFE) and codon adaptation index (CAI). LinearDesign represents the space of synonymous coding sequences as a deterministic finite automaton (DFA) and uses lattice parsing to optimize the mRNA sequence and its secondary structure simultaneously. For a protein sequence of length $|p|$, its joint objective is $\mathrm{MFE}-\lambda |p|\log(\mathrm{CAI})$, where the user-specified parameter $\lambda$ controls the relative importance of codon adaptation. Setting $\lambda=0$ yields MFE-only design, whereas increasing $\lambda$ places greater emphasis on host-preferred codon usage. 
The implementation employs beam pruning, retaining only a fixed number of candidate dynamic-programming states at each step. This improves scalability, although the resulting search is approximate and is not guaranteed to recover the global optimum of the unpruned objective. In contrast to our ensemble-based formulation, which marginalizes over the full Boltzmann ensemble and directly samples coding sequences from the resulting distribution, LinearDesign returns a single sequence optimized with respect to its predicted minimum-energy structure. 
While the published LinearDesign algorithm describes beam pruning and includes an algorithm for its implementation, the publicly available code does not allow for beam pruning.
(Note that benchmarks with beam pruning are provided for EnsembleDesign.)

\subsection*{EnsembleDesign} Next, EnsembleDesign\cite{Dai2025-EnsembleDesign} extends LinearDesign by replacing its deterministic codon lattice with a probabilistic lattice representing a distribution over valid synonymous sequences. Whereas LinearDesign optimizes the energy of a single minimum-free-energy structure, EnsembleDesign seeks favorable stability across the full Boltzmann ensemble of possible secondary structures. It applies probabilistic lattice parsing to calculate an expected partition function and uses projected gradient descent to optimize the lattice probabilities.
The publicly available code exposes beam pruning and other options, which we benchmark below.

\subsection*{CDSfold} Finally, CDSfold \cite{Terai2015} designs a synonymous protein-coding sequence with minimum predicted folding free energy. CDSfold extends the Zuker RNA-folding dynamic program to operate under amino-acid constraints, allowing codon selection and secondary-structure prediction to be performed simultaneously. The algorithm evaluates standard structural motifs, including stems, hairpins, bulges, internal loops, and multiloops, and traceback produces a single MFE-optimized coding sequence and its predicted structure. Its basic objective does not explicitly optimize codon adaptation, although selected codons may be excluded. The exact algorithm requires $O(n^3)$ time and $O(n^2)$ memory. Like LinearDesign, CDSfold optimizes a single minimum-energy structure rather than the full Boltzmann ensemble.

\subsection*{Performance results}

\Cref{tab:lineardesign_results,tab:cdsfold_results,tab:ensembledesign_results} summarize our computational benchmarks (respectively) for LinearDesign, EnsembleDesign, and CDSfold across example systems 1-4.
These runtimes may be compared to the optimized CPU-GPU runtimes yielded by our approach (cf. \cref{tab:runtime-table}).
For comparison, we list the wall-clock time to sample 2500 sequences via our approach (including partition function calculation and simultaneous sampling) in the right-hand column of each table.
In general, we find that our tensor-based approach is roughly 10-1000$\times$ faster in our benchmarks.
Input and output files for the compared benchmarks have been made publicly available at \url{https://github.com/mfornace/codon-optimization-benchmarks}.

In addition, unlike these alternative methods, our approach yields a full distribution which can be sampled repeatedly with minimal expense (\cref{tab:runtime-table}).
To be clear, note that the free energy models and objective differ substantially between the compared methods.
We therefore intend this comparison only to provide a rough guide for an evaluation of the speed and scalability of the different algorithms.

\begin{table}[H]
    \centering
    \caption{Performance benchmarks for our example systems using LinearDesign. Results were obtained with $\lambda = 0$ (MFE-only design) and $\lambda = 4$. 
    As mentioned above, beam pruning is not an allowed option in the publicly available source code.}
    \label{tab:lineardesign_results}
    \begin{tabular}{c|cc|c}
    \toprule
    & \multicolumn{2}{c|}{Alternative method runtime (s)} & Tensor-based runtime (s) \\
    \midrule
    $\lambda$ & 0 & 4 &  \\
    \midrule
    System &  &  &  \\
    \midrule
    1 & 15 & 15 & 0.7 \\
    2 & 129 & 129 & 5.0 \\
    3 & 529 & 524 & 18 \\
    4 & 3144 & 3170 & 81 \\
    \bottomrule
    \end{tabular}
\end{table}

\begin{table}[H]
    \centering
    \caption{Performance benchmarks for our example systems using EnsembleDesign.
    Results were obtained using a learning rate of 0.03, $\epsilon=0.5$, and 64 CPU threads. 
    Here, $\epsilon$ controls the trade-off between random initialization and MFE initialization, and the number of runs determines the number of output sequences.
    Results labeled ``---'' exceeded available memory ($\approx$ 512 GB).
    The result labeled ``$>$ 172800'' timed out at 48 hours due to cluster job policy.
    }
    \label{tab:ensembledesign_results}

    \begin{tabular}{c|cccccccc|c}
    \toprule
    & \multicolumn{8}{c|}{Alternative method runtime (s)} & Tensor-based runtime (s) \\
    \midrule
    Beam pruning & \multicolumn{4}{c|}{None} & \multicolumn{4}{c|}{200} &  \\
    Iterations & \multicolumn{2}{c|}{3} & \multicolumn{2}{c|}{30} & \multicolumn{2}{c|}{3} & \multicolumn{2}{c|}{30} &  \\
    Runs & \multicolumn{1}{c|}{1} & \multicolumn{1}{c|}{10} & \multicolumn{1}{c|}{1} & \multicolumn{1}{c|}{10} & \multicolumn{1}{c|}{1} & \multicolumn{1}{c|}{10} & \multicolumn{1}{c|}{1} & \multicolumn{1}{c|}{10} & \\
    \midrule
    System &  &  &  &  &  &  &  &  &  \\
    \midrule
    1 & 198 & 214 & 1815 & 1833 & 86 & 99 & 726 & 746 & 0.7 \\
    2 & 1842 & 2112 & 15986 & 16526 & 395 & 607 & 2581 & 2822 & 5.0 \\
    3 & 7926 & --- & 69323 & --- & 1186 & --- & 5841 & --- & 18 \\
    4 & 51949 & --- & $>$ 172800 & --- & 5227 & --- & 16952 & --- & 81 \\
    \bottomrule
    \end{tabular}
\end{table}

\begin{table}[H]
    \centering
    \caption{Performance benchmarks for our example systems using CDSFold. Default parameters were used.}
    \label{tab:cdsfold_results}
    \begin{tabular}{c|c|c}
    \toprule
    System & Alternative method runtime (s) & Tensor-based runtime (s) \\
    \midrule
    1 & 21 & 0.7 \\
    2 & 345 & 5.0 \\
    3 & 2658 & 18 \\
    4 & 20031 & 81 \\
    \bottomrule
    \end{tabular}
\end{table}

Each published code was run on the same CPU-based hardware (NERSC Perlmutter cluster, AMD EPYC 7763 processors, 512 GB RAM, up to 64 cores) as our own tensor-based implementation.
None of the compared codes provided GPU acceleration, and only EnsembleDesign exposed an option for parallelism.


%% file: pseudocode/pairs.tex
\begin{algorithm}
    \caption{$O(n^3)$ algorithm to calculate reverse intervals}
    \label{alg:pairs}
    \begin{algorithmic}
        \Function{ComputeReverse}{$Q, \Nphi, \DesignFold$}
            \For{$i \in \UpTo{\Nphi}$}
                \ForAll{$j \in \Range{i}{\Nphi}$} \Comment{Calculate folding of wraparound interval $\Range{j}{i}$}
                        \State $R_{i,j} \gets \DesignFold_{j, i+n}(Q_{j+1,i-1} \textbf { if } j+1 \leq i-1 \textbf{ else } R_{j+1,i-1})$ 
                \EndFor
                \ForAll{$j \in \Exclusive{i}{\Nphi}$}
                    \State $R_{j,i} \gets \sum_{k \in \Range{j}{\Nphi}} Q_{j,k} R_{i,k}^\t
                                        + \sum_{k \in \UpTo{i}} R_{j,k} Q_{i,k}^\t$
                \EndFor
            \EndFor
            \State \Return $R$
        \EndFunction
    \end{algorithmic}
\end{algorithm}

%% file: pseudocode/sample.tex
\begin{algorithm}
	\newcommand{\Alts}{\vec{v}}
	\newcommand{\Op}{z}
	\small
	\caption{Simultaneous Boltzmann sampling}
	\label{alg:sample}
	\begin{algorithmic}
		\Function{SampleWeights}{$\Ix, T$} \LComment{Associate random weights in $\RealRange{0}{T}$ to each element in index sequence $\Ix$ and return the resultant pairs in the order in which those weights are ascending}
			\State $\Ix' \gets \Call{UniformlyShuffled}{\Ix}$
			\State $T' \gets T$
			\For{$n = \Abs{\Ix}, \dots, 1$}
				\State $T' \gets T' \cdot (\Call{UniformRandom}{0, 1})^{1/n}$
				\State $w_{n-1} \gets (T', \Ix'_{n-1})$ \Comment{Each returned element has a weight and an index}
			\EndFor
			\State \Return $\vec{w}$
		\EndFunction
		\newline
		\Procedure{SampleAlternatives}{$T, \Ix, \Alts, \Queue, \PartialStrucs$} 
			\LComment{Simultaneously sample alternative states given total partition function $T$, state indices $\Ix$, alternatives $\Alts$, recursion element map $\Queue$, and state list $\PartialStrucs$}
			\State $\vec{w} \gets \Call{SampleWeights}{\UpTo{J}, T}$
			\State $t \gets 0$
			\For{$(\vec{E},\vec{\Op},t') \in \Alts$}
				\State $t \gets t + t'$
				\If{$\Abs{\vec{w}} > 0$ and $\Call{FirstWeight}{\vec{w}} < t$}
					\State $(\vec{u}, \vec{w}) \gets \Call{UpperBound}{\vec{w}, t}$ \Comment{Partition elements (already sorted) with weights $\leq t$ and $> t$}
					\ForAll{$E \in \vec{E}$}
						\State Insert recursion element $E$ into $\Queue$ if absent, concatenate indices of $\vec{u}$ to its mapped index list
					\EndFor
					\ForAll{$i \in \vec{u}, \Op \in \vec{\Op}$}
						\State Annotate states $s_i$ with operator index $\Op$
					\EndFor
				\EndIf
			\EndFor
		\EndProcedure
		\newline
		\Function{SampleStates}{$J, Q, \Nphi$} \Comment{$J$: the number of samples desired}
			\State $\Queue \gets \Call{EmptyPriorityQueue}{ }$ \Comment{Map from recursion element to the indices of associated states}
			\State $\PartialStrucs \gets$ list of $J$ empty states
			\newline
			\LComment{Sample from the unconstrained partition function $\TrS[Q_{0, \Nphi}]$}
			\State $\Alts \gets \lrb{\lrp{\lrb{E_{0,n}^{\sigma,a,\sigma,b}}, S^{b, a} Q_{0,n}^{a, b}}  \text{ for all } a \in \Dims, b \in \Dims, \sigma \in \UpTo{\Sigma_0}}$
			\State $\Call{SampleAlternatives}{\TrS[Q_{0, \Nphi}],\UpTo{J},v,\Queue,\PartialStrucs}$
			\newline
			\LComment{Sample from remaining recursion elements until none remain}
			\While{$\Abs{\Queue} > 0$} 
				\State $(E_{i,j}^{\sigma_a,\zeta_a,\sigma_b,\zeta_b}, \Ix) \gets \Call{Pop}{\Queue}$
				\ForAll{$k \in \Ix$}
					\State Annotate $s_k$ with unpaired base $j$ if $j+1=i$ or base pair $(j, i-1)$ if $j+1<i$
				\EndFor
				\newline
				\LComment{Determine all alternative states for this recursion element}
				\If{$i < j$}
					\State $\Alts \gets \lrb{\lrp{\lrb{E_{i,k}^{\sigma_a,\zeta_a,\sigma_c,\zeta_c},\; E_{j,k}^{\sigma_b, \zeta_b,\sigma_c,\zeta_c}},\; \emptyset, Q_{i,k}^{\sigma_a,\zeta_a,\sigma_c,\zeta_c} Q_{j,k}^{\sigma_b,\zeta_b,\sigma_c,\zeta_c}} \text{ for all } k \in \Range{i}{j}, \zeta_c \in \Dims, \sigma_c \in \UpTo{\Sigma_k}}$
				\ElsIf{$j+1 = i$}
					\State $\Alts \gets \lrb{\lrp{\emptyset, \lrb{z}, x \cdot \V_{\phi_j}^{\zeta_b,\zeta_a}} \text{ for all } (z, \sigma_j, \phi_j, \sigma_{j+1}, x) \in \Call{Operators}{j} \text{ s.t. } \sigma_b = \sigma_j \text{ and } \sigma_a = \sigma_{j+1}}$
				\ElsIf{$j < i$}
					\State $\Alts \gets \Big\{ \begin{aligned}[t] & \lrp{\lrb{E_{j+1,i-1}^{\sigma_{j+1},c,\sigma_{i-1},d}}, \lrb{z, z'}, 
						x \cdot x' \cdot \displaystyle\sum_{c \in \Dims,e \in \Dims} B_{\phi_j, \phi_{i-1}}^{\rho,\zeta_a,\zeta_b} S^{d,e} B_{\phi_{i-1}, \phi_j}^{\rho,e,c} Q_{j+1,i-1}^{\sigma_{j+1},c,\sigma_{i-1},d}} 
						\\ & \hspace{-2.5em}\text{ for all } (z,\sigma_j,\phi_j,\sigma_{j+1},x) \in \Call{Operators}{j}, \;
						                     (z',\sigma_{i-1},\phi_{i-1},\sigma_i,x') \in \Call{Operators}{i-1} 
											 \text{ s.t } \sigma_a = \sigma_i \text{ and } \sigma_b = \sigma_j
											 \Big\} \end{aligned} $
				\EndIf
				\newline
				\State $\Call{SampleAlternatives}{Q_{i,j}^{a,b}, \Ix, \Alts, \Queue, \PartialStrucs}$
			\EndWhile
		\EndFunction
		\State \Return $\vec{s}$
	\end{algorithmic}
\end{algorithm}

%% file: pseudocode/subopt.tex
\begin{algorithm}
	\newcommand{\Alts}{\vec{v}}
	\newcommand{\Op}{z}
	\newcommand{\qopt}{q^\mathrm{opt}}
	\newcommand{\Best}{E^\mathrm{opt}}
	\caption{Suboptimal state generation}
	\small
	\label{alg:subopt}
	\begin{algorithmic}
		\Procedure{ScanAlternatives}{$\qopt, \Ix, \Alts, \gap, \Queue, \PartialStrucs$} 
			\LComment{Consider alternative states given optimal value $\qopt$, state indices $\Ix$, alternatives $\Alts$, free energy gap $\gap$, recursion element map $\Queue$, and state list $\PartialStrucs$}
			\ForAll{$(\vec{E}, \vec{\Op}, q) \in \Alts$}
				\If{$q = \qopt$ and $\Best$ is unset} \Comment{Track first-found optimal state separately}
					\State{$\Best \gets \vec{E}$}
					\State{Annotate $s_k$ with each operator $\Op \in \vec{\Op}$}
				\ElsIf{$q \leq \qopt + \gap$}
					\For{$i \in \UpTo{\Call{LowerBound}{\Ix, \gap + q - \qopt}}$} \Comment{Find feasible structures by binary search}
						\State $(r, \vec{E'}, q') \gets s_{\Ix_i}$ 
						\State Append to $\vec{s}$ with partial state $r$, gap $q'+\qopt-q$, recursion indices $\vec{E} \cup \vec{E'}$, annotated by each operator $\Op \in \vec{\Op}$
						\For{$E \in \vec{E} \cup \vec{E'}$}
							\State Append new state index $\Abs{\PartialStrucs}-1$ to the mapped list $\Queue_E$
						\EndFor
					\EndFor
				\EndIf
			\EndFor
			\ForAll{$E \in \Best$}
				\State Append each of state indices $\Ix$ to the mapped list $\Queue_E$
				\State Append recursion element $E$ to the structures of indices $\Ix$
			\EndFor
		\EndProcedure
		\newline
		\Function{SuboptimalStates}{$\gap, Q, \Nphi$} 
			\State $\Queue \gets \Call{EmptyOrderedMap}{ }$ \Comment{Map from recursion element to list of associated state indices}
			\State $\PartialStrucs \gets \{(\Call{EmptyAnnotatedState}{\Nphi}, 0, \emptyset)\}$ \Comment{Initialize state list with an empty state of gap 0 and no recursion elements}
			\newline
			\LComment{Recurse through the trace $\TrS[Q_{0,\Nphi}]$}
			\State $\Alts \gets \lrb{(E_{0,n}^{\sigma,a,\sigma,b}, S^{b, a} + Q_{0,n}^{a, b})  \text{ for all } a \in \Dims, b \in \Dims, \sigma \in \UpTo{\Sigma_0}}$
			\State $\Call{ScanAlternatives}{\TrS[Q_{0,\Nphi}], \lrb{0}, \Alts, \gap, \Queue, \PartialStrucs}$
			\newline
			\LComment{Backtrack through recursion elements until none remain}
			\While{$\Abs{\Queue} > 0$} 
				\State $(E_{i,j}^{\sigma_a,\zeta_a,\sigma_b,\zeta_b}, \Ix) \gets \Call{Pop}{\Queue}$
				\State Sort partial state indices $\Ix$ by their associated free energy gap (ascending)
				\ForAll{$k \in \Ix$}
					\State Remove recursion element $E_{i,j}^{\sigma_a,\zeta_a,\sigma_b,\zeta_b}$ from state $s_k$
					\State Annotate $s_k$ with unpaired base $j$ if $j+1=i$ or base pair $(j, i-1)$ if $j+1<i$
				\EndFor
				\newline
				\LComment{Determine all alternative states for this recursion element}
				\If{$i < j$}
					\State $\Alts \gets \lrb{\lrp{\lrb{E_{i,k}^{\sigma_a,\zeta_a,\sigma_c,\zeta_c}, E_{j,k}^{\sigma_b, \zeta_b,\sigma_c,\zeta_c}}, \emptyset, Q_{i,k}^{\sigma_a,\zeta_a,\sigma_c,\zeta_c} + Q_{j,k}^{\sigma_b,\zeta_b,\sigma_c,\zeta_c}} \text{ for all } k \in \Range{i}{j}, \zeta_c \in \Dims, \sigma_c \in \UpTo{\Sigma_k}}$
				\ElsIf{$j+1 = i$}
					\State $\Alts \gets \lrb{\lrp{\emptyset, \lrb{z}, q + \V_{\phi_j}^{\zeta_b,\zeta_a}} \text{ for all } (z, \sigma_j, \phi_j, \sigma_{j+1}, q) \in \Call{Operators}{j} \text{ s.t. } \sigma_b = \sigma_j \text{ and } \sigma_a = \sigma_{j+1}}$
				\ElsIf{$j < i$}
					\State $\Alts \gets \Big\{ \begin{aligned}[t] & \lrp{\lrb{E_{j+1,i-1}^{\sigma_{j+1},c,\sigma_{i-1},d}}, \lrb{z, z'}, 
						q + q' + \displaystyle\min_{c \in \Dims,e \in \Dims} B_{\phi_j, \phi_{i-1}}^{\rho,\zeta_a,\zeta_b} + S^{d,e} + B_{\phi_{i-1}, \phi_j}^{\rho,e,c} + Q_{j+1,i-1}^{\sigma_{j+1},c,\sigma_{i-1},d}} 
						\\ &\hspace{-2.5em}\text{ for all } (z,\sigma_j,\phi_j,\sigma_{j+1},q) \in \Call{Operators}{j}, \;
						                     (z',\sigma_{i-1},\phi_{i-1},\sigma_i,q') \in \Call{Operators}{i-1} 
											 \text{ s.t } \sigma_a = \sigma_i \text{ and } \sigma_b = \sigma_j
											 \Big\} \end{aligned} $
				\EndIf
				\newline
				\State $\Call{ScanAlternatives}{Q_{i,j}^{a,b}, \Ix, \Alts, \gap, \Queue, \PartialStrucs}$
			\EndWhile
			\State \Return $\PartialStrucs$
		\EndFunction
	\end{algorithmic}
\end{algorithm}

%% file: floats/frequency-matching.tex
\begin{figure}[H]
    \centering
    \includegraphics[width=0.8\columnwidth]{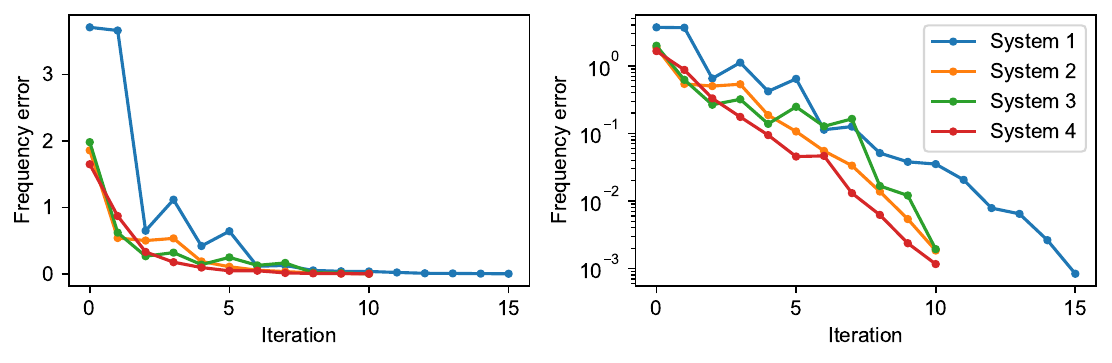} 
       \caption{\label{fig:frequency-matching}
            Demonstration of codon frequency matching via fixed point iteration with Anderson acceleration (with no history size limit).
            Single codon frequencies were fit for each of benchmark systems 1-4 (\cref{si:study-details}), with the target frequencies matching those used to construct CAI bonuses.
            Calculations were run in single precision using up to two GPUs.
            \Bullet{a} Maximum absolute logarithmic error ($\Norm{\log(b/b_*)}_\infty$) as a function of iteration.
            Iterations were terminated once this error was below $r_\mathrm{tol} = 0.002$.
            \Bullet{b} Same as (a), with $y$-values plotted logarithmically.
       }
\end{figure}

%% file: tables/codon-frequencies-a.tex
\begin{tabular}[t]{cccc}
        \begin{tabular}[t]{llr}
\toprule
 Amino acid & Codon & Frequency  \\
\midrule
\multirow[t]{4}{*}{A} & GCU & 0.122 \\
 & GCC & 0.408 \\
 & GCA & 0.077 \\
 & GCG & 0.392 \\
\cline{1-3}
\multirow[t]{6}{*}{R} & CGU & 0.029 \\
 & CGC & 0.291 \\
 & CGA & 0.056 \\
 & CGG & 0.457 \\
 & AGA & 0.039 \\
 & AGG & 0.128 \\
\cline{1-3}
\multirow[t]{2}{*}{N} & AAU & 0.449 \\
 & AAC & 0.551 \\
\cline{1-3}
\multirow[t]{2}{*}{D} & GAU & 0.552 \\
 & GAC & 0.448 \\
\cline{1-3}
\multirow[t]{2}{*}{C} & UGU & 0.410 \\
 & UGC & 0.590 \\
\cline{1-3}
\multirow[t]{2}{*}{Q} & CAA & 0.151 \\
 & CAG & 0.849 \\
\cline{1-3}
\multirow[t]{2}{*}{E} & GAA & 0.167 \\
 & GAG & 0.833 \\
\cline{1-3}
G & GGU & 0.079 \\
\cline{1-3}
\end{tabular}
 & 
\begin{tabular}[t]{llr}
\toprule
 Amino acid & Codon & Frequency  \\
\midrule
\multirow[t]{3}{*}{G} & GGC & 0.370 \\
 & GGA & 0.107 \\
 & GGG & 0.443 \\
\cline{1-3}
\multirow[t]{2}{*}{H} & CAU & 0.727 \\
 & CAC & 0.273 \\
\cline{1-3}
\multirow[t]{3}{*}{I} & AUU & 0.212 \\
 & AUC & 0.631 \\
 & AUA & 0.157 \\
\cline{1-3}
\multirow[t]{6}{*}{L} & CUU & 0.075 \\
 & CUC & 0.287 \\
 & CUA & 0.029 \\
 & CUG & 0.495 \\
 & UUA & 0.020 \\
 & UUG & 0.093 \\
\cline{1-3}
\multirow[t]{2}{*}{K} & AAA & 0.085 \\
 & AAG & 0.915 \\
\cline{1-3}
M & AUG & 1. \\
\cline{1-3}
\multirow[t]{2}{*}{F} & UUU & 0.224 \\
 & UUC & 0.776 \\
\cline{1-3}
\multirow[t]{2}{*}{P} & CCU & 0.070 \\
 & CCC & 0.476 \\
\cline{1-3}
\end{tabular}
 & 
\begin{tabular}[t]{llr}
\toprule
 Amino acid & Codon & Frequency  \\
\midrule
\multirow[t]{2}{*}{P} & CCA & 0.077 \\
 & CCG & 0.377 \\
\cline{1-3}
\multirow[t]{6}{*}{S} & UCU & 0.054 \\
 & UCC & 0.288 \\
 & UCA & 0.060 \\
 & UCG & 0.113 \\
 & AGU & 0.161 \\
 & AGC & 0.323 \\
\cline{1-3}
\multirow[t]{4}{*}{T} & ACU & 0.239 \\
 & ACC & 0.392 \\
 & ACA & 0.137 \\
 & ACG & 0.232 \\
\cline{1-3}
W & UGG & 1. \\
\cline{1-3}
\multirow[t]{2}{*}{Y} & UAU & 0.129 \\
 & UAC & 0.871 \\
\cline{1-3}
\multirow[t]{4}{*}{V} & GUU & 0.093 \\
 & GUC & 0.499 \\
 & GUA & 0.044 \\
 & GUG & 0.364 \\
\cline{1-3}
\end{tabular}

    \end{tabular}
    

%% file: tables/codon-frequencies-b.tex
\begin{tabular}[t]{cccc}
        \begin{tabular}[t]{llr}
\toprule
 Amino acid & Codon & Frequency  \\
\midrule
\multirow[t]{4}{*}{A} & GCU & 0.306 \\
 & GCC & 0.228 \\
 & GCA & 0.110 \\
 & GCG & 0.356 \\
\cline{1-3}
\multirow[t]{6}{*}{R} & CGU & 0.163 \\
 & CGC & 0.140 \\
 & CGA & 0.118 \\
 & CGG & 0.381 \\
 & AGA & 0.032 \\
 & AGG & 0.167 \\
\cline{1-3}
\multirow[t]{2}{*}{N} & AAU & 0.418 \\
 & AAC & 0.582 \\
\cline{1-3}
\multirow[t]{2}{*}{D} & GAU & 0.431 \\
 & GAC & 0.569 \\
\cline{1-3}
\multirow[t]{2}{*}{C} & UGU & 0.484 \\
 & UGC & 0.516 \\
\cline{1-3}
\multirow[t]{2}{*}{Q} & CAA & 0.142 \\
 & CAG & 0.858 \\
\cline{1-3}
\multirow[t]{2}{*}{E} & GAA & 0.198 \\
 & GAG & 0.802 \\
\cline{1-3}
G & GGU & 0.286 \\
\cline{1-3}
\end{tabular}
 & 
\begin{tabular}[t]{llr}
\toprule
 Amino acid & Codon & Frequency  \\
\midrule
\multirow[t]{3}{*}{G} & GGC & 0.193 \\
 & GGA & 0.101 \\
 & GGG & 0.419 \\
\cline{1-3}
\multirow[t]{2}{*}{H} & CAU & 0.562 \\
 & CAC & 0.438 \\
\cline{1-3}
\multirow[t]{3}{*}{I} & AUU & 0.309 \\
 & AUC & 0.513 \\
 & AUA & 0.179 \\
\cline{1-3}
\multirow[t]{6}{*}{L} & CUU & 0.132 \\
 & CUC & 0.234 \\
 & CUA & 0.055 \\
 & CUG & 0.379 \\
 & UUA & 0.053 \\
 & UUG & 0.148 \\
\cline{1-3}
\multirow[t]{2}{*}{K} & AAA & 0.169 \\
 & AAG & 0.831 \\
\cline{1-3}
M & AUG & 1. \\
\cline{1-3}
\multirow[t]{2}{*}{F} & UUU & 0.282 \\
 & UUC & 0.718 \\
\cline{1-3}
\multirow[t]{2}{*}{P} & CCU & 0.242 \\
 & CCC & 0.232 \\
\cline{1-3}
\end{tabular}
 & 
\begin{tabular}[t]{llr}
\toprule
 Amino acid & Codon & Frequency  \\
\midrule
\multirow[t]{2}{*}{P} & CCA & 0.085 \\
 & CCG & 0.441 \\
\cline{1-3}
\multirow[t]{6}{*}{S} & UCU & 0.101 \\
 & UCC & 0.172 \\
 & UCA & 0.071 \\
 & UCG & 0.361 \\
 & AGU & 0.071 \\
 & AGC & 0.224 \\
\cline{1-3}
\multirow[t]{4}{*}{T} & ACU & 0.206 \\
 & ACC & 0.230 \\
 & ACA & 0.083 \\
 & ACG & 0.481 \\
\cline{1-3}
W & UGG & 1. \\
\cline{1-3}
\multirow[t]{2}{*}{Y} & UAU & 0.557 \\
 & UAC & 0.443 \\
\cline{1-3}
\multirow[t]{4}{*}{V} & GUU & 0.210 \\
 & GUC & 0.382 \\
 & GUA & 0.083 \\
 & GUG & 0.326 \\
\cline{1-3}
\end{tabular}

    \end{tabular}
    

%% file: tables/codon-frequencies-c.tex
\begin{tabular}[t]{cccc}
        \begin{tabular}[t]{llr}
\toprule
 Amino acid & Codon & Frequency  \\
\midrule
\multirow[t]{4}{*}{A} & GCU & 0.236 \\
 & GCC & 0.227 \\
 & GCA & 0.091 \\
 & GCG & 0.446 \\
\cline{1-3}
\multirow[t]{6}{*}{R} & CGU & 0.139 \\
 & CGC & 0.206 \\
 & CGA & 0.088 \\
 & CGG & 0.348 \\
 & AGA & 0.028 \\
 & AGG & 0.191 \\
\cline{1-3}
\multirow[t]{2}{*}{N} & AAU & 0.443 \\
 & AAC & 0.557 \\
\cline{1-3}
\multirow[t]{2}{*}{D} & GAU & 0.449 \\
 & GAC & 0.551 \\
\cline{1-3}
\multirow[t]{2}{*}{C} & UGU & 0.328 \\
 & UGC & 0.672 \\
\cline{1-3}
\multirow[t]{2}{*}{Q} & CAA & 0.178 \\
 & CAG & 0.822 \\
\cline{1-3}
\multirow[t]{2}{*}{E} & GAA & 0.151 \\
 & GAG & 0.849 \\
\cline{1-3}
G & GGU & 0.140 \\
\cline{1-3}
\end{tabular}
 & 
\begin{tabular}[t]{llr}
\toprule
 Amino acid & Codon & Frequency  \\
\midrule
\multirow[t]{3}{*}{G} & GGC & 0.273 \\
 & GGA & 0.117 \\
 & GGG & 0.470 \\
\cline{1-3}
\multirow[t]{2}{*}{H} & CAU & 0.327 \\
 & CAC & 0.673 \\
\cline{1-3}
\multirow[t]{3}{*}{I} & AUU & 0.248 \\
 & AUC & 0.520 \\
 & AUA & 0.232 \\
\cline{1-3}
\multirow[t]{6}{*}{L} & CUU & 0.118 \\
 & CUC & 0.217 \\
 & CUA & 0.060 \\
 & CUG & 0.342 \\
 & UUA & 0.037 \\
 & UUG & 0.226 \\
\cline{1-3}
\multirow[t]{2}{*}{K} & AAA & 0.189 \\
 & AAG & 0.811 \\
\cline{1-3}
M & AUG & 1. \\
\cline{1-3}
\multirow[t]{2}{*}{F} & UUU & 0.355 \\
 & UUC & 0.645 \\
\cline{1-3}
\multirow[t]{2}{*}{P} & CCU & 0.151 \\
 & CCC & 0.290 \\
\cline{1-3}
\end{tabular}
 & 
\begin{tabular}[t]{llr}
\toprule
 Amino acid & Codon & Frequency  \\
\midrule
\multirow[t]{2}{*}{P} & CCA & 0.078 \\
 & CCG & 0.481 \\
\cline{1-3}
\multirow[t]{6}{*}{S} & UCU & 0.093 \\
 & UCC & 0.174 \\
 & UCA & 0.043 \\
 & UCG & 0.261 \\
 & AGU & 0.167 \\
 & AGC & 0.261 \\
\cline{1-3}
\multirow[t]{4}{*}{T} & ACU & 0.216 \\
 & ACC & 0.247 \\
 & ACA & 0.128 \\
 & ACG & 0.409 \\
\cline{1-3}
W & UGG & 1. \\
\cline{1-3}
\multirow[t]{2}{*}{Y} & UAU & 0.417 \\
 & UAC & 0.583 \\
\cline{1-3}
\multirow[t]{4}{*}{V} & GUU & 0.171 \\
 & GUC & 0.298 \\
 & GUA & 0.099 \\
 & GUG & 0.431 \\
\cline{1-3}
\end{tabular}

    \end{tabular}
    

%% file: tables/codon-frequencies-d.tex
\begin{tabular}[t]{cccc}
        \begin{tabular}[t]{llr}
\toprule
 Amino acid & Codon & Frequency  \\
\midrule
\multirow[t]{4}{*}{A} & GCU & 0.128 \\
 & GCC & 0.363 \\
 & GCA & 0.093 \\
 & GCG & 0.416 \\
\cline{1-3}
\multirow[t]{6}{*}{R} & CGU & 0.135 \\
 & CGC & 0.308 \\
 & CGA & 0.076 \\
 & CGG & 0.332 \\
 & AGA & 0.031 \\
 & AGG & 0.118 \\
\cline{1-3}
\multirow[t]{2}{*}{N} & AAU & 0.334 \\
 & AAC & 0.666 \\
\cline{1-3}
\multirow[t]{2}{*}{D} & GAU & 0.339 \\
 & GAC & 0.661 \\
\cline{1-3}
\multirow[t]{2}{*}{C} & UGU & 0.243 \\
 & UGC & 0.757 \\
\cline{1-3}
\multirow[t]{2}{*}{Q} & CAA & 0.142 \\
 & CAG & 0.858 \\
\cline{1-3}
\multirow[t]{2}{*}{E} & GAA & 0.170 \\
 & GAG & 0.830 \\
\cline{1-3}
G & GGU & 0.161 \\
\cline{1-3}
\end{tabular}
 & 
\begin{tabular}[t]{llr}
\toprule
 Amino acid & Codon & Frequency  \\
\midrule
\multirow[t]{3}{*}{G} & GGC & 0.381 \\
 & GGA & 0.079 \\
 & GGG & 0.379 \\
\cline{1-3}
\multirow[t]{2}{*}{H} & CAU & 0.364 \\
 & CAC & 0.636 \\
\cline{1-3}
\multirow[t]{3}{*}{I} & AUU & 0.180 \\
 & AUC & 0.566 \\
 & AUA & 0.255 \\
\cline{1-3}
\multirow[t]{6}{*}{L} & CUU & 0.077 \\
 & CUC & 0.276 \\
 & CUA & 0.062 \\
 & CUG & 0.410 \\
 & UUA & 0.027 \\
 & UUG & 0.148 \\
\cline{1-3}
\multirow[t]{2}{*}{K} & AAA & 0.142 \\
 & AAG & 0.858 \\
\cline{1-3}
M & AUG & 1. \\
\cline{1-3}
\multirow[t]{2}{*}{F} & UUU & 0.296 \\
 & UUC & 0.704 \\
\cline{1-3}
\multirow[t]{2}{*}{P} & CCU & 0.124 \\
 & CCC & 0.346 \\
\cline{1-3}
\end{tabular}
 & 
\begin{tabular}[t]{llr}
\toprule
 Amino acid & Codon & Frequency  \\
\midrule
\multirow[t]{2}{*}{P} & CCA & 0.116 \\
 & CCG & 0.414 \\
\cline{1-3}
\multirow[t]{6}{*}{S} & UCU & 0.063 \\
 & UCC & 0.218 \\
 & UCA & 0.073 \\
 & UCG & 0.250 \\
 & AGU & 0.117 \\
 & AGC & 0.280 \\
\cline{1-3}
\multirow[t]{4}{*}{T} & ACU & 0.123 \\
 & ACC & 0.346 \\
 & ACA & 0.114 \\
 & ACG & 0.417 \\
\cline{1-3}
W & UGG & 1. \\
\cline{1-3}
\multirow[t]{2}{*}{Y} & UAU & 0.491 \\
 & UAC & 0.509 \\
\cline{1-3}
\multirow[t]{4}{*}{V} & GUU & 0.091 \\
 & GUC & 0.387 \\
 & GUA & 0.075 \\
 & GUG & 0.447 \\
\cline{1-3}
\end{tabular}

    \end{tabular}
    